# Optimized Look-Ahead Tree Policies:
# A Bridge Between Look-Ahead Tree Policies and Direct Policy Search

**Tobias Jung**[1]
(corresponding author)
Phone:+32 4 366 5605
tjung@ulg.ac.be

**Louis Wehenkel**[1]
L.Wehenkel@ulg.ac.be

**Damien Ernst**[1]
dernst@ulg.ac.be

**Francis Maes**[1]
francis.maes@ulg.ac.be

[1]Montefiore Institute
Department of Electrical Engineering and Computer Science
University of Liège
Belgium

# Optimized Look-Ahead Tree Policies: A Bridge Between Look-Ahead Tree Policies and Direct Policy Search

**Abstract**

Direct policy search (DPS) and look-ahead tree (LT) policies are two widely used classes of techniques to produce high performance policies for sequential decision-making problems. To make DPS approaches work well, one crucial issue is to select an appropriate space of parameterized policies with respect to the targeted problem. A fundamental issue in LT approaches is that, to take good decisions, such policies must develop very large look-ahead trees which may require excessive online computational resources. In this paper, we propose a new hybrid policy learning scheme that lies at the intersection of DPS and LT, in which the policy is an algorithm that develops a small look-ahead tree in a directed way, guided by a node scoring function that is learned through DPS. The LT-based representation is shown to be a versatile way of representing policies in a DPS scheme, while at the same time, DPS enables to significantly reduce the size of the look-ahead trees that are required to take high-quality decisions.

We experimentally compare our method with two other state-of-the-art DPS techniques and four common LT policies on four benchmark domains and show that it combines the advantages of the two techniques from which it originates. In particular, we show that our method: (1) produces overall better performing policies than both pure DPS and pure LT policies, (2) requires a substantially smaller number of policy evaluations than other DPS techniques, (3) is easy to tune and (4) results in policies that are quite robust with respect to perturbations of the initial conditions.

**Short title:** Optimized look-ahead tree policies

# 1 Introduction

A wide range of techniques have been developed in the last decades for solving optimal sequential decision-making problems in the fields of optimal control and reinforcement learning. Two important classes of techniques that are known to work well on difficult problems characterized by large state spaces are *direct policy search* (DPS) [4] and *look-ahead tree* (LT) policies [23].

In DPS, a policy is seen as a parameterized function that maps states to actions. In order to identify the best settings of the parameters, DPS techniques rely on local or global optimization algorithms that directly maximize the performance of the policy estimated through simulation or real-world experiments. DPS avoids learning value functions and all the complexities involved with approximating them. Instead, optimization is carried out directly in policy space. The rationale for doing this is that, while usually the value function will be a rather complex object and thus thought to be hard to approximate, a control policy will often be a much "simpler" object and thus could also be (so it is believed) much easier to learn. Note that in DPS approaches, making a decision is generally a very fast operation while all heavy computations are performed in an offline way, when optimizing the parameters of the policy.

Look-ahead tree policies rely on a totally different principle than DPS, since they have no parameters that need to be learned in an offline way: all the computational work is done online during each decision-making step. In order to take each single decision, a LT policy proceeds in two steps: first, it develops a look-ahead tree by simulating the evolution of the system subject to multiple possible future action sequences, then, it selects an action based on the information collected in this tree. The quality of decisions taken by LT policies depends on the size of the tree: the more action sequences have been tried, the better the final decision will be. The computational complexity of an LT policy can thus be adjusted by changing the size of the look-ahead tree, small trees leading to fast approximate decisions and large trees leading to expensive high-quality decisions.

Both DPS and LT techniques have their strengths and weaknesses. DPS can often produce, with comparatively little effort, good policies for difficult control problems where typical value function based methods would otherwise fail. However, a major issue in DPS is that the final performance strongly depends on the choice of an appropriate policy representation. Common policy representations include linear parametrizations [22], neural networks [37, 47, 19] or radial basis functions [6, 14] and typically have hyper-parameters that require tuning (e.g. the number of hidden neurons). For a given problem, choosing an appropriate representation and tuning its hyper-parameters is a difficult problem, which typically involves a large amount of trial and error. On the other hand, this effort is not required with LT policies thanks to the look-ahead tree that acts as a unified built-in highly generic representation of the decision problem. Hence, LT policies can be applied to almost any sequential decision-making problem without any prior knowledge – this without even depending on the dimensionality of the state space. The main issue with LT policies is that they typically require a large amount of online computational resources to produce high-quality decisions. In practice, these resources are often limited, either due to real-time constraints (e.g. 100 ms per decision) or human-scale constraints (e.g. one hour or one day per decision). Given such a constraint and given the difficulty of the problem, the actions suggested by LT policies may be arbitrarily sub-optimal.

LT policies alleviates the need to choose a complex approximation structure for policies, while at the same time DPS offers a principled way of reducing online computational requirements through an offline learning procedure. Starting from two these observations, it is natural to wonder how we can combine these two ideas in order to create a new hybrid solution for sequential decision-making that leverages the advantages of both.

In this paper, we propose to bridge the gap between LT and DPS policies with a new hybrid technique: *optimized look-ahead tree (OLT) policies*[1]. From the point of view of LT policies, our approach starts from the observation that the efficiency of an LT policy, given a finite computational budget, can be improved by directing the way in which the tree is grown, for example by giving priority to sequences of actions that seem more promising [23]. In fact, a large number of different strategies could be considered to develop look-ahead trees, by combining aspects from breadth-first exploration to depth-first exploration and from random exploration to purely greedy exploration. In practice, the quality of these different strategies depend on the characteristics of the problem (e.g. to which extent are rewards informative about the long-term goal?), on the available online computational budget, and may even also depend on the current location in the state-space. Optimized look-ahead trees then consist in parameterizing a node scoring function that encodes the way in which look-ahead trees are grown and then learning the best values of these parameters in a problem-driven way through DPS.

---
[1]This technique was initially proposed in [32] and is here extended through a more mature exposition of the method and an extensive experimental study which covers much more aspects than the initial paper. Furthermore, we introduce the use of Gaussian process optimization to improve the sample efficiency of the approach.

Traditionally, policies in DPS rely on an approximation structure, which is usually one of those used in supervised learning. These approximation structures lead to policies which are representable through simple mathematical formulas, such as the maximization of a linear dot-product or the computation of a feed-forward neural network. Note however that, choosing such a policy or an OLT policy does not make any conceptual difference for DPS: as any other policy, OLT policies may be viewed as functions that receive a state, perform some parameter-dependent internal computations and then return an action. The distinguishing feature of OLT policies is that their internal computations rely on explicitly exploiting a generative model of the problem.

We experimentally compare our hybrid method with two other state-of-the-art DPS techniques, where the policies are represented through neural networks and adaptive radial basis functions, respectively. We also consider LT policies with four different exploration strategies. Based on an extensive study on four challenging benchmark domains, we show that our approach indeed leverages the advantages of the two techniques from which it originates. In particular, we show first that OLT policies perform often significantly better than both pure DPS and pure LT policies and that they are at least competitive in all cases. Second, thanks to their use of a model, it turns out that OLT policies require a substantially smaller number of parameters than neural networks and radial basis functions to reach high-performance policies. In turn, they require, as shown in the experimental results, significantly less offline policy evaluations than other DPS techniques. Third, we show that OLT policies are much more straightforward to train as compared to usual DPS which often requires cumbersome trial and error iterations to choose and train their approximation structure. Finally, we show that OLT policies are quite robust with respect to perturbations of the initial conditions.

The content of the paper is structured as follows. In Section 2, we begin by introducing basic notation and state formally the type of sequential decision-making problems this paper is about. Section 3 presents a detailed description of optimized look-ahead trees. The node scoring function of optimized look-ahead trees can be learned using any derivative-free global optimizer. One such optimizer which was shown to be highly relevant to DPS is Gaussian process optimization [17, 29]. We give a brief (but for all practical purposes fully sufficient) summary of this approach in Section 4. Section 5 presents the results of extensive experimental evaluations, wherein we compare optimized look-ahead trees with different pure DPS and pure LT techniques on a number of well-known benchmark domains. We present related work in Section 6 and finally conclude in Section 7.

## 2 Problem statement

This section introduces basic notation and formally states the problem we are going to solve. We consider a deterministic discrete-time system whose dynamics are given by $x_{t+1} = f(x_t, a_t)$, where $x$ is an element of some state space $X$ and $a$ is an element of some action space $A$. We do not make any assumptions about the form of the state space, since the presented approach can deal with any form of $X$. We do, however, make strong assumptions about the action space $A$: we require that it is finite, i.e. $|A| = K$, and, for computational reasons, that the number of actions is small. In the examples we will consider later, $X$ will be continuous and vector-valued, i.e., $X \subset \mathbb{R}^{n_x}$ and each discrete action will be mapped to a specific continuous control vector (bang-bang control).

The system evolves at discrete time steps with $t = 1, 2, \ldots$; for every transition we make, we observe a scalar *reward* $\varrho(x_t, a_t)$ which serves as a performance measure; we denote by $\underline{B} = \inf\{\varrho(x_t, a_t) : (x, a) \in X \times A\}$ and $\overline{B} = \sup\{\varrho(x_t, a_t) : (x, a) \in X \times A\}$, without however assuming that they are finite. Let $\pi : X \to A$ denote a stationary *policy*, i.e. a mapping from states to actions. For any given policy $\pi$ and state $x_0$, the *infinite horizon discounted sum of rewards* $V^\pi(x_0)$ is defined as

$$V^\pi(x_0) := \lim_{T \to \infty} \sum_{t=0}^{T} \gamma^t \varrho(x_t, \pi(x_t)), \quad \text{where } \forall t: \ x_{t+1} = f(x_t, \pi(x_t)), \tag{1}$$

and where $0 < \gamma < 1$ is a discount factor. The optimal value function is defined as the maximum over all policies, $V^*(x_0) := \max_\pi V^\pi(x_0)$, and satisfies the discrete-time Hamilton-Jacobi-Bellman (HJB) equation

$$V^*(x) = \max_a \left[\varrho(x, a) + \gamma V^*\big(f(x, a)\big)\right] \quad \forall x \in X. \tag{2}$$

If we are able to determine the solution $V^*$ of Eq. (2), the optimal action for every state $x$ can be derived as $\pi^*(x) = \mathrm{argmax}_a \left[ \varrho(x,a) + \gamma V^*\bigl(f(x,a)\bigr) \right]$. However, solving Eq. (2) exactly is possible only in some cases (e.g., LQR); in the general nonlinear case and for higher dimensional spaces this presents an open research problem.

DPS approaches do not try to estimate $V^\pi$ or $V^*$. Instead, they solve the conceptually simpler problem of finding a policy that "works well" for some initial conditions. In our case, we define this objective as follows: given a set of initial states $X_0 \subset X$, our goal is to find a policy that maximizes the performance over the states[2] in $X_0$, i.e., we want to find

$$\mathrm{argmax}_\pi \sum_{x_0 \in X_0} V^\pi(x_0) \tag{3}$$

(or $\mathrm{argmax}_\pi \sum_{x_0 \in X_0} p(x_0) V^\pi(x_0)$, where $p(x_0) \geq 0$ are some weights).

In order to solve this problem, one assumes that policies are functions $\pi_\theta := \pi(\cdot\,;\theta)$ parameterized by some vector $\theta \in \mathbb{R}^d$. The optimization over policies in Eq. (3) is thus turned into an optimization over real-valued vectors

$$\mathrm{argmax}_{\theta \in \mathbb{R}^d} V_{X_0}(\theta) := \sum_{x_0 \in X_0} V^{\pi_\theta}(x_0). \tag{4}$$

Given a vector $\theta$, the objective function $V_{X_0}$ can be evaluated by simulating the system under the policy $\pi_\theta$ from all the states in $X_0$ and summing the rewards as in Eq. (1). Note that, to avoid the infinite sum, we have to truncate the infinite horizon to a (typically large) number of finite simulation steps $H$. The objective function in Eq. (4) is thus effectively replaced by

$$\mathrm{argmax}_{\theta \in \mathbb{R}^d} V_{X_0}(\theta) := \sum_{x_0 \in X_0} \sum_{t=0}^{H} \gamma^t \varrho\bigl(x_t, \pi(x_t;\theta)\bigr), \quad \text{where } \forall t: \; x_{t+1} = f(x_t, \pi(x_t;\theta)). \tag{5}$$

The novelty of this paper is that we consider policies $\pi(\cdot\,;\theta)$ which, instead of being defined through function approximators, are non-trivial algorithms that at some point depend on $\theta$ and on the generative model $(f, \varrho)$. To emphasize this requirement, we will write $\pi_{f,\varrho}(\cdot\,;\theta)$ whenever we refer explicitly to an OLT policy.

## 3 Optimized Look-Ahead Trees

Given a state $x_t$, we now describe a way of selecting action $\pi_{f,\varrho}(x_t;\theta)$ based on the construction of a look-ahead tree. The construction of the tree will, in general, be non-uniform and is controlled by a function that is parameterized by $\theta$ (non-uniform meaning that leaf nodes are not necessarily at the same depth).

### 3.1 Notation

Let $K$ be the number of possible actions and $\mathcal{T}$ denote the look-ahead tree. $\mathcal{T}$ is composed of nodes $n_{i,h} \in \mathcal{T}$, where $h$ denotes the depth and $i$ denotes the index within the depth $h$ (i.e., $i \in \{1, \ldots, K^h\}$). The state $x_t$ for which we want to find action $\pi_{f,\varrho}(x_t;\theta)$ is placed in the root node $n_{1,0}$. Each node $n_{i,h}$ corresponds to a particular sequence of actions and states. The children of $n_{i,h}$ are generated by applying one of the $K$ actions: assume $a^{(h)} \in \{1, \ldots, K\}$ is the index of the action taken at depth $h$, then the child of $n_{i,h}$ is $n_{K(i-1)+a^{(h)}, h+1}$.

For each node $n_{i,h}$ there exists one unique path from the root $n_{1,0}$ to $n_{i,h}$:

$$\text{path from } n_{1,0} \text{ to } n_{i,h}: \quad n_{1,0} \longrightarrow n_{i_1,1} \longrightarrow n_{i_2,2} \longrightarrow \ldots \longrightarrow n_{i_h,h} = n_{i,h}$$

which corresponds to a particular sequence of actions $a^{(1)}, a^{(2)}, \ldots, a^{(h)}$ taken at each intermediate step and which induces a partial trajectory of length $h$:

$$x_t \xrightarrow[\varrho(x_t, a^{(1)})]{a^{(1)}} x_{t+1}^{a^{(1)}} \xrightarrow[\varrho(x_{t+1}^{a^{(1)}}, a^{(2)})]{a^{(2)}} x_{t+2}^{a^{(1)} a^{(2)}} \ldots x_{t+h-1}^{a^{(1)} \ldots a^{(h-1)}} \xrightarrow[\varrho(x_{t+h-1}^{a^{(1)} \ldots a^{(h-1)}}, a^{(h)})]{a^{(h)}} x_{t+h}^{a^{(1)} \ldots a^{(h)}}$$

---

[2]Note that in the experiments we report on later we will only use a single initial state. In the general case, having multiple initial states tends to increase the robustness of the policy.

The successor states are generated according to the transition function $f$, e.g., $x_{t+i}^{a^{(1)}\ldots a^{(i)}} = f\left(x_{t+i-1}^{a^{(1)}\ldots a^{(i-1)}}, a^{(i)}\right)$ and rewards according to the reward function $\varrho$. For each node $n_{i,h}$ we define $\varrho(n_{i,h})$ to be the reward obtained in the last step: $\varrho(n_{i,h}) := \varrho(x_{t+h-1}^{a^{(1)}\ldots a^{(h-1)}}, a^{(h)})$.

Every time we expand a node, we generate all of its children. A node whose children have been generated is called an *inner node*. Otherwise it is called a *terminal node*[3]. We denote $\mathcal{T} = \mathcal{T}_{\text{inner}} \cup \mathcal{T}_{\text{leaf}}$.

## 3.2 Using look-ahead trees to make decisions

We first describe how to use the information contained in a look-ahead tree to select an action. For each terminal node $n_{i,h} \in \mathcal{T}_{\text{leaf}}$ we define the $\ell$-*score* as the discounted sum of rewards obtained along the path from the root node $n_{1,0}$ to $n_{i,h}$ plus a lower bound on the cumulated rewards not yet observed:

$$\forall n_{i,h} \in \mathcal{T}_{\text{leaf}}: \quad \ell(n_{i,h}) := \sum_{t=1}^{h} \varrho(n_{i_t,t})\gamma^{t-1} + \sum_{t=h+1}^{\infty} \underline{B}\gamma^{t-1} = \sum_{t=1}^{h} \varrho(n_{i_t,t})\gamma^{t-1} + \frac{\underline{B}\gamma^h}{1-\gamma}. \quad (6)$$

For each non-terminal node $n_{i,h} \in \mathcal{T}_{\text{inner}}$ we define the $\ell$-score recursively as the maximum of the $\ell$-scores of its children (see Figure 1):

$$\forall n_{i,h} \in \mathcal{T}_{\text{inner}}: \quad \ell(n_{i,h}) := \max_{n \in \text{children}(n_{i,h})} \ell(n). \quad (7)$$

The $\ell$-score of the root node $n_{1,0}$ (which corresponds to state $x_t$) is a lower bound on the optimal value $V^*(x_t)$ [23]. Given a look-ahead tree, we adopt the conservative strategy that consists in selecting the action that leads to the successor state with maximal lower bound. With the naming scheme for nodes introduced above, we can write $\ell(n_{1,0}) = \max_a \ell(n_{a,1})$, and thus

$$\pi_{f,\varrho}(x_t) = \underset{a}{\arg\max}\ \ell(n_{a,1}). \quad (8)$$

In the rest of this paper, we generally assume that $\underline{B} = 0$ and that $\overline{B}$ is finite, although neither of these conditions is a requirement for our OLT method.

## 3.3 Developing look-ahead trees

We now discuss the construction of look-ahead trees. If the rewards are upper-bounded ($\overline{B}$ is finite), we can also upper bound the optimal values $V^*(x)$. For each terminal node $n_{i,h} \in \mathcal{T}_{\text{leaf}}$ we define the $u$-*score* as the discounted sum of rewards obtained along the path from the root node $n_{1,0}$ to $n_{i,h}$, plus an upper bound on the cumulated rewards not yet observed:

$$\forall n_{i,h} \in \mathcal{T}_{\text{leaf}}: \quad u(n_{i,h}) := \sum_{t=1}^{h} \varrho(n_{i_t,t})\gamma^{t-1} + \sum_{t=h+1}^{\infty} \overline{B}\gamma^{t-1} = \ell(n_{i,h}) + \frac{(\overline{B} - \underline{B})\gamma^h}{1-\gamma}. \quad (9)$$

For each non-terminal node $n_{i,h} \in \mathcal{T}_{\text{inner}}$ we define the $u$-score recursively as the maximum of the $u$-scores of its children (see Figure 1):

$$\forall n_{i,h} \in \mathcal{T}_{\text{inner}}: \quad u(n_{i,h}) := \max_{n \in \text{children}(n_{i,h})} u(n). \quad (10)$$

See Figure 1 for a graphical illustration of a look-ahead tree with associated $\ell$-scores and $u$-scores. The $u$-score of the root node $n_{1,0}$ is an upper bound on the optimal value $V^*(x_t)$. Furthermore, the $\ell$-score of the root increases with the number of expanded nodes in the tree, while at the same time its $u$-score decreases [23]. In other words, the more nodes the tree has and the farther we develop it into the future, the tighter our bounds on the optimal value will become[4] and thus the better a decision we can hope to make with Eq. (8). (Bear in mind

---

[3] Note that the following terminology is equivalent:

non-terminal node = inner node = explored node = closed node
terminal node = leaf node = unexplored node = open node.

[4] The tightness of the bounds will also depend on $\gamma$. The closer $\gamma$ is to 1, the more nodes we will have to expand. However, there is little we can do about it, since $\gamma$ is given in advance.

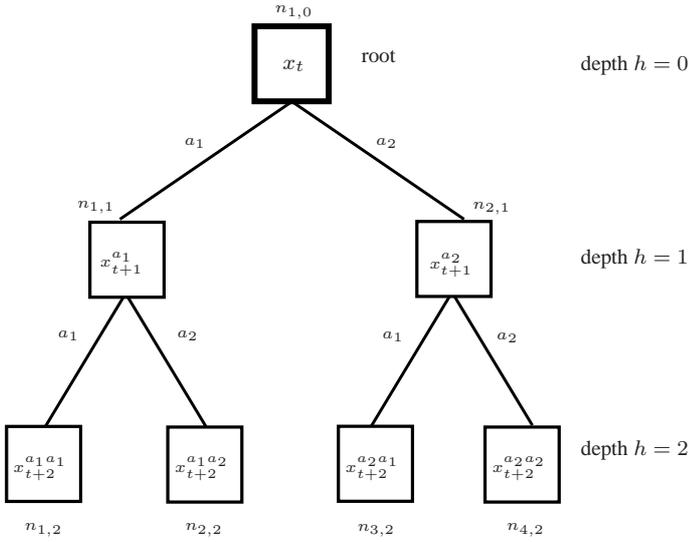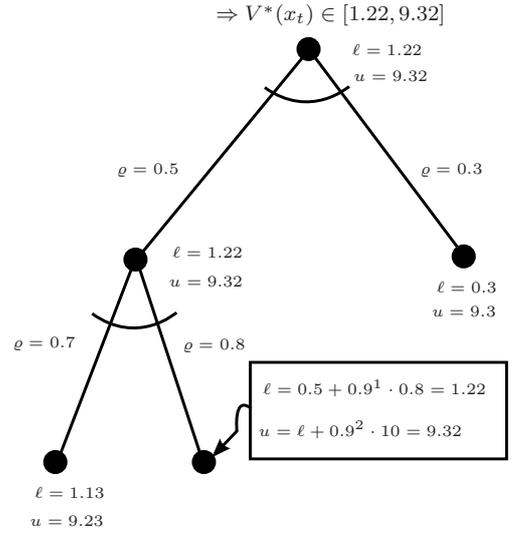

Figure 1: Left side: shows a uniform look-ahead tree developed from state $x_t$ for $h = 2$ stages and $K = 2$ actions. In it, nodes are labeled by $n_{i,h}$ and correspond to states; edges correspond to choosing an action. Each node encodes the result of taking a particular sequence of actions; e.g., the node $n_{2,2}$ corresponds to state $x_{t+2}^{a_1 a_2}$, which is the result of starting from state $x_t$ and choosing action $a_1$ in the first stage and action $a_2$ in the second stage. Right side: illustrates how $\ell$-scores and $u$-scores are calculated (which form an upper and lower bound for the true value $V^*(x_t)$ of the root node). The plot shows a non-uniform tree where the terminal node with the highest $u$-score is expanded first. Edges are labeled with the rewards $\varrho$ associated with the transition they represent. Note that the $\ell$-scores and $u$-scores are calculated only for terminal nodes and propagated back to their parent via the *max* operator (denoted by the arc).

that, unlike search trees for constraint satisfaction problems which are grown up to a goal state, our look-ahead trees are grown until a predefined computational budget is exhausted.)

Knowing that, the big question is: *how should we develop the tree such that we arrive at a near-optimal decision within the allowed computational budget?*

The typical way of developing a look-ahead tree is to build a uniform tree, by expanding nodes with a breadth-first strategy. However, to fully develop a tree of depth $h$ we need $K^0 + K^1 + K^2 + \ldots + K^h$ node expansions which for growing $h$ will rapidly become computationally infeasible.

Best-first search is the common solution to this dilemma; it develops trees non-uniformly and only expands those nodes to a deeper depth that look "promising". In order to do this, the algorithm relies on a *node scoring function* $e : \mathcal{T}_{\text{leaf}} \to \mathbb{R}$ that is used to assign a certain score to every terminal node $n_{i,h} \in \mathcal{T}_{\text{leaf}}$; whenever we want to decide which terminal node to expand next, we compare all the scores and choose the node with the highest score. Note that the default uniform development strategy can be obtained as a particular case of best-first search with the following node scoring function:

$$e^{uniform}(n_{i,h}) := -h. \tag{11}$$

The work in [23] suggests to use as scoring function the $u$-scores of the terminal nodes (which makes the method become closely related to A*):

$$e^{optimistic}(n_{i,h}) := u(n_{i,h}). \tag{12}$$

The authors of [23] show theoretically that non-uniform trees developed by the $u$-score will never perform worse than uniform trees for the same budget of expansions, but certain conditions must be met to make them perform better. The extent to which these conditions are fulfilled is problem-specific and depends on how "informative" the reward is (which in turn depends on the coupling of dynamics and rewards). Informally speaking, a problem will be difficult (i.e., the rewards will be non-informative) if there are many nodes with the property that, given the observed rewards from the root to the node in question, one cannot decide whether the node lies on an optimal path or not. This is, for example, the case for reinforcement learning problems with a flat reward structure, such as in the mountain car domain (where every transition has 0 reward and only entering the goal gives +1). On the other hand, having an informative reward is not a totally exotic requirement; for many control problems an informative reward comes as the natural definition of performance (e.g., the many pole balancing or inverted pendulum domains, where rewards are taken as a quadratic function of angle and angular velocity).

### 3.4 Parameterizing the development of the look-ahead tree

It turns out that a large number of strategies could be used to develop look-ahead trees and it is probably the case that there exists no single best strategy for all problems. Instead of searching for the best possible generic strategy, we adopt the approach first introduced in [32], which consists in learning a specific look-ahead node development strategy in a problem-driven way.

Optimized look-ahead trees rely on a parameterized node scoring function $e(\cdot\,;\theta) : \mathcal{T}_{\text{leaf}} \to \mathbb{R}$, where $\theta \in \mathbb{R}^d$ is a parameter vector. This function should be flexible enough to represent a large variety of tree development strategies. In particular the parameters should encode important aspects of search, such as the extent to which depth-first search is preferred over breadth-first search, or how short-term rewards should be used to bias search. Furthermore, since different search strategies may be optimal in different regions of the state space, we would like $e(\cdot\,;\theta)$ to enable state-dependent strategies. To meet these requirements, as in [32], we take the parameterized node scoring function to be a simple weighted sum of features extracted from the information encoded in the path from the root node to the node in question.

In our investigations, the node scoring function is defined for each terminal node $n_{i,h} \in \mathcal{T}_{\text{leaf}}$ in the following way: let $x_{t+h}^{a^{(1)}\ldots a^{(h)}} = (x_{t+h}^{(1)}, \ldots, x_{t+h}^{(n_x)}) \in \mathbb{R}^{n_x}$ denote the $n_x$-dimensional state corresponding to $n_{i,h}$, and $\varrho_{t+h} = \varrho(n_{i,h})$ denote its reward. We consider three blocks of features: the first $n_x$ features merely correspond to the components of the state, the next $n_x$ features correspond to components of the state multiplied by the reward (enabling the tree to be grown in a more or less directed way, and the final set of $n_x$ features corresponds to the components of the state multiplied by the depth $h$ (enabling to control breadth/depth trade-off). Let $\theta \in \mathbb{R}^{3n_x}$ be the vector of parameters. The parameterized scoring function $e(\cdot\,;\theta)$ can then be written as

$$\forall n_{i,h} \in \mathcal{T}_{\text{leaf}} : \quad e(n_{i,h};\theta) := \sum_{j=1}^{n_x} x_{t+h}^{(j)} (\theta_j + \theta_{n_x+j} \cdot \varrho_{t+h} + \theta_{2n_x+j} \cdot h). \tag{13}$$

Notice that with this linear parameterization the outcome of which node is selected is invariant under scaling of the parameter vector by positive scalars. Of course, other kinds of features are also possible and may in fact turn out to be more suitable in some cases (this may be an avenue for future research).

### 3.5 Connection with the optimal value function

If in place of scoring function $e(\cdot\,;\theta)$ we would have used the optimal value function $V^*$ to decide which node to expand next, the algorithm in Eq. (8) would have returned the optimal action for any number of node expansions $\geq 1$. In this sense, scoring function $e(\cdot\,;\theta)$ can be seen as a weaker form of value function. It is weaker because it does not have to assign the precise quantity "discounted future sum of rewards"; any number will do as long as it helps to grow the tree in roughly the right direction.

It is for this reason that we believe it could be advantageous to try to learn a good scoring function instead of trying to directly learn/approximate the optimal value function: the former can be a rather simple function (as evidenced in the good results we get in Section 5 where we use for all domains exactly the same simple weighted sum of features), whereas the latter would require a far more expressive parametrization (and it is well-known that value function approximation scales badly when the dimensionality of the state space grows [41]).

### 3.6 Summary: the algorithm

Figure 2 presents a simple algorithm based on a sorted list to implement policies as parameterized look-ahead trees. The algorithm requires as input a state $x_t$ and the parameter vector $\theta$ and returns the action $\pi_{f,\varrho}(x_t;\theta)$. It

**Input:** state $x_t$, weights $\theta$
**Output:** policy action $\pi_{f,\varrho}(x_t;\theta)$
Depends on:
    $f$    transition function
    $\varrho$    reward function
    $\gamma$    discount factor
    $e(\cdot;\theta)$    parameterized score function
    $h_{\max}$    maximum number of node expansions

Node $n$ is a `struct` consisting of fields:
    $n.x$    state
    $n.h$    depth
    $n.\ell$    $\ell$-score=cumulative reward
    $n.e$    tree development score from $e(\cdot;\theta)$
    $n.\pi$    first action on path

**1. Initialize**
    list=$\emptyset$
    for $i = 1\ldots K$ /* Expand root node for all actions */
        $x' := f(x_t, a_i)$; $r' := \varrho(x_t, a_i)$; $h' := 1$; $e := e(x', r', h';\theta)$
        generate node $n$:
            $n.x := x'$; $n.\ell := r'$; $n.h := h'$; $n.e := e$; $n.\pi := a_i$
        add $n$ to list

**2. Main loop**
    While number of node expansions $j < h_{\max}$ (the max number of node expansions)
        Find $n^* := \mathrm{argmax}_{n \in list}\, n.e$ and remove $n^*$ from list
        for $i = 1\ldots K$ /* Expand node $n^*$ for all actions */
            $x' := f(n^*.x, a_i)$; $r' := \varrho(n^*.x, a_i)$; $h' := n^*.h + 1$; $e := e(x', r', h';\theta)$
            generate node $n$:
                $n.x := x'$; $n.\ell := n^*.\ell + \gamma^{h'-1} r'$; $n.h := h'$; $n.e := e$; $n.\pi := n^*.\pi$
            add $n$ to list

**3. Get best action**
    return $\pi_{f,\varrho}(x_t;\theta) := n^*.\pi$ where $n^* = \mathrm{argmax}_{n \in list}\, n.\ell$ /* ties broken randomly */

Figure 2: Implementing a policy represented by a parameterized look-ahead tree.

depends on the allowed number of node expansions $h_{max}$, the domain (represented by generative model $f$, $\varrho$) and the discount factor $\gamma$. The computational complexity for evaluating $\pi_{f,\varrho}(x_t;\theta)$ for a budget of $h_{max}$ node expansions is obtained as follows. Let $D$ be the cost for one call to the generative model and $E$ be the cost to recursively update the internal scores of a node. We assume that the leaf nodes are stored in an appropriate data structure which allows incremental insertion in logarithmic time and allows finding the maximimum of the scores in constant time (such that for every iteration of the main loop, if the the list of terminal nodes contains $N$ elements, we can find the maximum score and update the structure with $\log N$ operations). Note that for every iteration $j = 1, 2, \ldots$ of the main loop, our list of terminal nodes contains $(K-1)(j-1) + K$ elements, which can be shown by simple induction (every time we expand a node we remove one element from the list and add $K$ new ones). At each iteration $j$ of the main loop we do the following: we first find the best node to expand by looping over the list of current terminal nodes. We then generate all of its $K$ successors, where each one costs $D$ for having to call the generative model and $E$ to recursively update its internal scores. After the main loop has run for $h_{\max}$ times, we have to find the best $\ell$-score from the list. The computational complexity of the algorithm is thus

$$\sum_{j=1}^{h_{\max}} \big[K(D+E) + \log\{(K-1)(j-1)+K\}\big] + (K-1)(h_{\max}-1) + K \leq$$

$$h_{\max}\big(K(D+E) + \log\{(K-1)(h_{\max}-1)+K\} + K - 1\big) + 1. \quad (14)$$

In the particular case where we are expanding by the heuristic *uniform* (i.e., breadth-first), we skip the maximization within each iteration of the main loop. The computational complexity then becomes

$$\sum_{j=1}^{h_{\max}} \big[K(D+E)\big] + (K-1)(h_{\max}-1) + K = h_{\max}\big(K(D+E) + K - 1\big) + 1. \quad (15)$$

Note that in order to expand uniformly, $h_{\max}$ must be equal to one of $K^0, K^0 + K^1, K^0 + K^1 + \ldots K^d$ where $d$ is the depth of the tree.

## 4 Gaussian Process Optimization

We now turn to the problem of solving Eq. (5), i.e., find a vector $\theta$ such that the induced policy $\pi(\cdot; \theta)$ (globally) maximizes the score $V_{X_0}(\theta)$ over the set of initial states $X_0$. In general, this will be a difficult optimization problem. First, the dependence of $V_{X_0}$ on $\theta$ can be a complex one with local extrema occuring frequently; thus we may need to sample the search space exhaustively. Second, there is no closed-form expression for evaluating the objective function $V_{X_0}$ or its gradient; instead we have to simulate the system (or run real-world experiments), which is expensive. Among the many alternatives that have been proposed in the past for this purpose, such as cross-entropy [44], various stochastic search alternatives [20], or Lipschitzian optimization [39], Gaussian process optimization (GPO) is considered to be one of the most efficient methods to optimize expensive functions [3].

The purpose of this section is to provide the required background in GPO to the reader who is not aware of this. Note that in our experiments we will illustrate the optimization of look-ahead trees with both GPO and the cross-entropy method, the latter being much simpler to understand and to implement than the former. Since choosing a policy representation and optimizing its parameters are two separate tasks, we keep the description of GPO general and independent of the specific nature of the OLT policies. Readers less interested in the optimization part may safely skip this section and directly jump to Section 5.

### 4.1 Notation

To enhance the readability of this section and to conform with the standard notation used in the literature, we will define a local notation for this section which overlaps with what we use in the remainder of the article. Specifically, we will now write $f(x)$ for the objective function we want to maximize (instead of $V_{X_0}(\theta)$) and $x$ for its input arguments (instead of $\theta$).

### 4.2 Overview

GPO is an iterative sampling-based search procedure which constructs a surrogate for the objective function and optimizes that in place of the original one. GPO is able to incorporate prior knowledge about the problem and provides a principled way to build and exploit the surrogate so as to trade off exploration and exploitation of the search space.

Each iteration of GPO consists of two steps. Assume that at iteration $n$ we have already sampled the objective function at locations $x_1, \ldots, x_n$ with values $f(x_1), \ldots, f(x_n)$. The first step is to fit a regression model using Gaussian process regression to the samples gathered so far, that is, to fit a regression model to the training data $\mathcal{D}_n := \{(x_i, f(x_i))\}_{i=1}^n$. Let us call the resulting model $GP_n$. The second step is to use the regression model $GP_n$ as input to a scoring heuristic $\mathcal{U}$ (the so-called *acquisition function*) to find the most promising point $x_{n+1}$ at which to evaluate the objective function next. Typically, acquisition functions are defined in an optimistic way such that high scores correspond to *potentially* high values of the objective function.

The regression model $GP_n$ acts as a surrogate of the objective function: it can be evaluated at any given point $x$ of the search space to produce an estimate for $f(x)$ (more precisely, $GP_n$ will produce a distribution over $f(x)$). The reason for building the surrogate is that, unlike the true objective function, it is computationally very cheap to evaluate (it can be done analytically and in closed form). Thus we can afford sampling it as exhaustively as necessary to find the maximum. On the other hand, since the values the surrogate produces will only be estimates, we can never be sure that what we get from maximizing the surrogate is indeed the maximum of the objective function or even close to it. Instead, we have to take into account how accurate the estimates produced by $GP_n$ are; this in turn will depend on the general smoothness of the objective function and the number of data points we have collected in the neighborhood of $x$.

Each time $GP_n$ is evaluated at a location $x$, it outputs two values $\mu_n(x)$ and $\sigma_n^2(x)$ which together define the Gaussian predictive distribution $N(\mu_n(x), \sigma_n^2(x))$ over values $f(x)$. The mean of this distribution, $\mu_n(x)$, can be directly taken as point estimate for $f(x)$. The variance of this distribution, $\sigma_n^2(x)$, can be taken as a measure

---

**Input:** objective function $V_{X_0}(\cdot)$ taking arguments $\theta \in \mathbb{R}^d$ (from Eq. (4))
**Output:** $\theta^* \approx \operatorname{argmax}_\theta V_{X_0}(\theta)$

**Depends on:**
   Acquisition function $\mathcal{U}$ that takes as functional input a GP model and as argument a vector $\theta \in \mathbb{R}^d$

---

**1. Initialize**
   Generate initial sample locations $\theta_1, \ldots, \theta_n$ by random sampling or space filling methods
      (e.g., Latin hypercube sampling)
   Evaluate the objective function in $\theta_1, \ldots, \theta_n$ to produce training data $\mathcal{D}_n := \{(\theta_i, V_{X_0}(\theta_i))\}_{i=1}^n$

**2. Main loop** $i = n, n+1, \ldots$ /* loop until we run out of computational resources */
   Fit GP model to the current training data:
      $GP_i$ = Gaussian process regression on training data $\mathcal{D}_i$
   Get most promising next sample location:
      $\theta_{i+1} := \operatorname{argmax}_\theta (\mathcal{U} \, GP_i)(\theta)$ /* using e.g. DIRECT */
   Evaluate objective function $V_{X_0}$ in $\theta_{i+1}$, add result to training data, and repeat:
      $\mathcal{D}_{i+1} = \mathcal{D}_i \cup \{(\theta_{i+1}, V_{X_0}(\theta_{i+1}))\}$

**3. Return**
   Training point in $\mathcal{D}_n$ with highest score:
      $\operatorname{argmax}_{\theta \in \mathcal{D}_n} V_{X_0}(\theta)$

---

Figure 3: Implementing Gaussian process optimization.

of how certain the GP is about this estimate. Both $\mu_n(x)$ and $\sigma_n^2(x)$ will be used by the acquisition function $\mathcal{U}$ to assign a score to $x$, which we will write as $(\mathcal{U} \, GP_n)(x)$. To determine the next sample location $x_{n+1}$, we thus have to determine

$$x_{n+1} := \operatorname*{argmax}_x (\mathcal{U} \, GP_n)(x) \tag{16}$$

which, unlike Eq. (5), can be solved efficiently by any black-box global optimization method (in our experimental studies we will use DIRECT [39]).

In the following two sections we will describe each of these steps in more detail; a summary of the algorithm is also given as pseudo-code in Figure 3.

### 4.3 Using Gaussian processes for regression

Here we briefly review how GPs can be used for function estimation [42]. Note that we will present GPs for the more general case of a stochastic objective function; while the problems we consider later are all deterministic, this leaves the door open for stochastic returns in future work.

Suppose we are looking for a function $f : \mathcal{X} \subset \mathbb{R}^d \to \mathbb{R}$ from which we have observed noisy samples $(x_1, y_1), \ldots, (x_n, y_n)$, where $x_i \in \mathcal{X}$ is the input and $y_i = f(x_i) + \varepsilon_i$ the output corrupted by independent zero-mean Gaussian noise with common variance $\sigma_0^2$, i.e., $\varepsilon_i \sim_{iid} \mathcal{N}(0, \sigma_0^2)$. To estimate the function value $f(x)$ at any given input location $x$, we proceed as follows. We suppose that the sought function is a realization of a zero-mean[5] Gaussian process with covariance function $k_\vartheta(x, x')$, which we write as $f \sim \mathcal{GP}(0, k_\vartheta(x, x'))$, where $\vartheta$ is a vector of hyper-parameters (as explained below). Hence, the vector of function values at the $n$ observed input locations is assumed to be drawn from a joint Gaussian distribution

$$(f(x_1), \ldots, f(x_n)) \mid X, \vartheta \sim \mathcal{N}(0_{n \times 1}, K_\vartheta), \tag{17}$$

where $X := [x_1, \ldots, x_n]$, $K_\vartheta$ is the $n \times n$ covariance matrix with entries $[K_\vartheta]_{i,j} = k_\vartheta(x_i, x_j)$. The covariance function $k_\vartheta(\cdot, \cdot)$ can be thought of as a way to encode our prior information about the "smoothness" of the functions $f$ we believe to come up; typically $k_\vartheta(x, x')$ is chosen as a function of the distance $\|x - x'\|$ in which case $k_\vartheta$ measures the expected amount of variation of the function values $f(x), f(x')$ in terms of the distance between the locations $x$ and $x'$.

---
[5]The assumption of zero-mean is made here for notational convenience. (Centering the data combined with an ergodicity assumption allows to avoid a non-zero mean.)

A typical choice for $k$ is the translation-invariant isotropic squared exponential, which is of the form

$$k_\vartheta(x, x') := v_0 \exp\{-0.5(x-x')^T \Omega (x-x')\} \qquad (18)$$

and which itself is parameterized by hyper-parameters $v_0 > 0$ and $\Omega = \text{diag}(a_1, \ldots, a_d)$, $a_i > 0$. All hyper-parameters specifying the GP are thus collected in the vector $\vartheta := (v_0, a_1, \ldots, a_d)$ which together with $\sigma_0^2$ characterizes the prior distribution of our noisy samples. The actual "training" of a GP thus consists of finding a good pair $(\vartheta, \sigma_0^2)$ from the data, for which various alternative procedures exist in the literature: here we stick to the most common one which is the optimization of the marginal likelihood. See [42] for a detailed description.

Now suppose that we know $\vartheta$ and $\sigma_0^2$. Since the noise is zero mean Gaussian, white, and independent of the function $f$, it follows from Eq. (17) that the $n \times 1$ vector of observed outputs $Y := [y_1, \ldots, y_n]^T$ will also be jointly Gaussian and distributed as follows

$$Y \mid X, \vartheta, \sigma_0^2 \sim \mathcal{N}(0_{n \times 1}, K_\vartheta + \sigma_0^2 I_{n \times n}). \qquad (19)$$

Furthermore, the joint distribution of the function value $f(x)$ at query location $x$ and the vector of observed values of $Y$ is also Gaussian and characterized in the following way:

$$\begin{bmatrix} Y \\ f(x) \end{bmatrix} \mid X, \sigma_0^2, \vartheta, x \sim \mathcal{N}\left(\begin{bmatrix} 0_{n \times 1} \\ 0 \end{bmatrix}, \begin{bmatrix} K_\vartheta + \sigma_0^2 I_{n \times n} & k_\vartheta(x) \\ k_\vartheta(x)^T & \kappa_\vartheta \end{bmatrix}\right), \qquad (20)$$

where the $n \times 1$ vector $k_\vartheta(x)$ is defined by $k_\vartheta(x) := [k_\vartheta(x, x_1), \ldots, k_\vartheta(x, x_n)]^T$ and scalar $\kappa_\vartheta$ by $\kappa_\vartheta := k_\vartheta(x, x)$. Conditioning $f(x)$ on $Y$, we thus obtain

$$f(x) \mid X, \vartheta, \sigma_0^2, x, Y \sim \mathcal{N}\left(\mu(x), \sigma^2(x)\right), \qquad (21)$$

where

$$\mu(x) := k_\vartheta(x)^T \left(K_\vartheta + \sigma_0^2 I_{n \times n}\right)^{-1} Y \qquad (22)$$

$$\sigma^2(x) := \kappa_\vartheta - k_\vartheta(x)^T \left(K_\vartheta + \sigma_0^2 I_{n \times n}\right)^{-1} k_\vartheta(x). \qquad (23)$$

Thus given (noisy) observations $(x_1, y_1), \ldots, (x_n, y_n)$ from an unknown function $f$, with GP regression we first infer from the observations hyper-parameters $\vartheta$ and $\sigma_0^2$, and then obtain for any new point $x$ the distribution over function values $p(f(x)|X, \vartheta, x, y) = \mathcal{N}(\mu(x), \sigma^2(x))$.

### 4.4 Choosing an acquisition function

Early work [24, 35] suggested to take as acquisition function the probability of improving over the current maximum $x^+ := \text{argmax}_{x \in \mathcal{D}_n} f(x)$ within the training data $\mathcal{D}_n$. The resulting PI acquisition function is given by $P(f(x) \geq f(x^+) + \zeta)$, which we write as

$$(PI\, GP_n)(x) := \Phi_{0,1}\left(\frac{\mu_n(x) - f(x^+) - \zeta}{\sigma_n(x)}\right) \qquad (24)$$

where $\mu_n(x)$ is the mean and $\sigma_n^2(x)$ the variance of the predictive distribution as output by $GP_n$ for point $x$ (see Eq. (22) and Eq. (23)), and $\Phi_{0,1}$ is the standard normal cumulative distribution. The trade-off parameter $\zeta \geq 0$ controls the compromise between the strength of the potential improvement and its probability to be realized.

An alternative acquisition function is the expected improvement EI, which can be evaluated analytically [24], giving

$$(EI\, GP_n)(x) := \sigma_n(x)\left(Z \Phi_{0,1}(Z) + \phi_{0,1}(Z)\right) \qquad (25)$$

where $Z := (\mu_n(x) - f(x^+) - \zeta)/\sigma_n(x)$, and $\phi_{0,1}$ is the probability density of the standard normal distribution.

Figure 4 illustrates how GPO works and how PI and EI give rise to distinct sampling behavior using the same GP and data. Which of the acquisition functions works best for a given problem is usually difficult to say in general and needs experimentation; in our examples in Section 5 we have found that EI worked best.

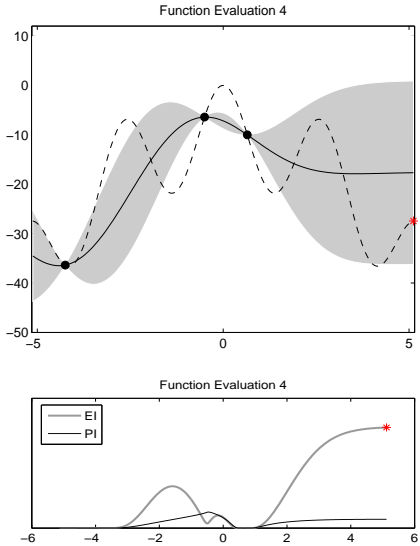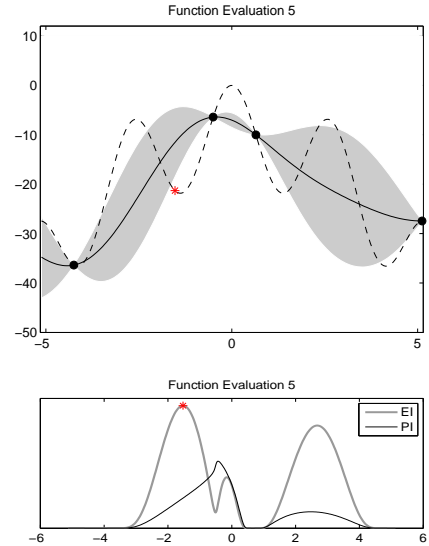

(a) $n = 4$ (after 3 function evaluations)

(b) $n = 5$ (after 4 function evaluations)

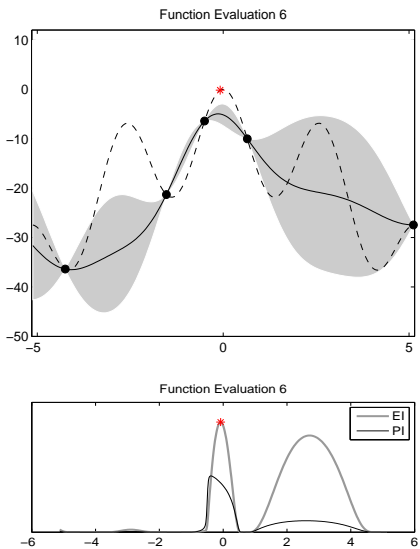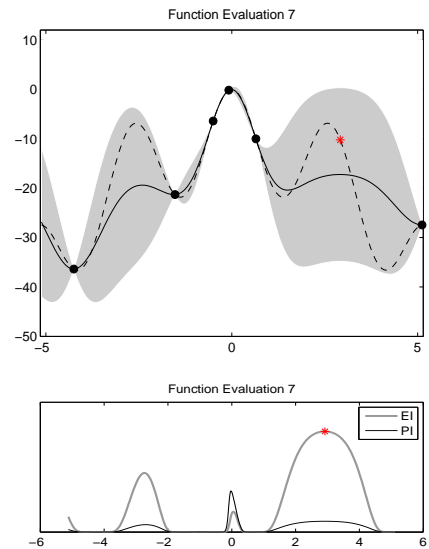

(c) $n = 6$ (after 5 function evaluations)

(d) $n = 7$ (after 6 function evaluations)

Figure 4: A concrete example illustrating Gaussian process optimization over four stages $n = 4, 5, 6, 7$. Each panel consists of two subplots. Each upper subplot shows the true (unknown) function we want to optimize as a dashed curve and the locations at which it was previously evaluated as black filled dots. Using these as training data $\mathcal{D}_n$ in a GP, the black curve denotes the expected value (mean) of the resulting predictive distribution from Eq. (22) and the shaded area denotes its variance from Eq. (23), both evaluated at all locations $x \in [-5, 5]$. As one would expect, the predictive variance (and thus the uncertainty about the estimates) is close to zero at "known" locations (it would be equal to zero if $\sigma_0^2 = 0$) and grows the farther one gets away from them. Each lower subplot shows the result of evaluating the acquisition function for one of the two possible choices EI or PI. Note that each of these would lead to a (slightly) different sampling behavior; here we have chosen the next sample location (marked in the figure by the star-shaped symbol) as the one that maximizes EI.

## 5 Experiments

This section reports an extensive evaluation of optimized look-ahead trees. The purpose of the experiments is to compare this approach to both pure DPS and pure LT approaches and to demonstrate its strengths across a variety of domains. To this end we have chosen four challenging benchmark problems, which are described in Section

5.1: inverted pendulum, double inverted pendulum, acrobot handstand and HIV drug treatment. We compare our approach to two of today's standard methods for DPS and to four generic LT policies, all of which are described in Section 5.2. The comparison is structured as follows:

- In Section 5.3, we compare the OLT approach to both state-of-the-art DPS approaches and LT approaches and show that OLT policies often significantly outperform these techniques and that they are at least competitive in all cases.

- In Section 5.4, we study the sample efficiency of OLT learning according to both the number of policy evaluations and the number of simulated transitions. We show that most of the time OLT requires less samples than its DPS competitors to reach high-performance policies.

- In Section 5.5, we compare how the different policy parameterization behave w.r.t. their respective hyper-parameters and show that optimized look-ahead trees require much less trial and error effort than the other studied DPS approaches.

- Finally, in Section 5.6, we study the robustness of LT, standard DPS, and OLT when the initial states of the testing set differ from the initial states of the training set and show that OLT policies are quite robust w.r.t. to both small and large perturbations of the initial state.

## 5.1 The benchmark domains

We consider four well-known challenging benchmark domains that exhibit some common characteristics: a continuous vector-valued state space, finite (discretized) actions, deterministic transitions and rewards that to some extent are informative about the goal. Each domain corresponds to a real physical process (either mechanical or bio-chemical) internally described by a system of nonlinear differential equations, none of which can be solved by traditional control methods such as LQR. The transition function for reinforcement learning is obtained by discretizing in time and keeping the controls constant; the actions are obtained by discretizing the bounded control space.

**Inverted pendulum** Our first domain is the inverted pendulum, a simple enough toy problem that is widely used in benchmarking different algorithms. The goal is to swing up and stabilize a single-link inverted pendulum as is shown in Figure 5a. As the motor does not provide enough torque to push the pendulum up in one single rotation, the pendulum needs first to be swung back and forth to gather energy before then being pushed up and balanced. This creates a nonlinear control problem. The state space is 2-dimensional, $x = (x_1, x_2) \equiv (\phi, \dot{\phi})$, with $\phi \in [-\pi, \pi]$ being the angle, and $\dot{\phi} \in [-10, 10]$ being the angular velocity. The control force is discretized to $a \in \{-5, -2.5, 0, +2.5, +5\}$ and held constant for $\Delta t = 0.2$sec. The dynamics of the system and physical parameters we used to instantiate the problem are specified in Appendix A.

To formulate this task as a reinforcement learning problem of the form given in Eq. (5), we used the following settings: the set of initial states is the singleton $X_0 = \{(\pi, 0)\}$, the reward is defined as $\varrho(x(t), u(t)) := 1 - 0.1(\phi(t)^2 - 0.1\dot{\phi}(t)^2 - 0.1u(t)^2)$, the discount factor is set to $\gamma = 0.99$, and each policy is evaluated for $H = 500$ steps.

**Double inverted pendulum** The next domain is a more complex variant of the inverted pendulum given above. This time we have two poles mounted each to a separate cart as depicted in Figure 5b. The two carts are linked by a spring and allowed to move some distance horizontally on the x-axis (until they collide with a wall which leads to a failure). Each cart is controlled separately; however, because of the spring their dynamics is coupled. As in the inverted pendulum, the goal is to swing up and stabilize the poles as quickly as possible, but now by moving the carts back and forth. This creates a rather challenging nonlinear control problem. The state space is 8-dimensional, $x \equiv (x_1, x_2, \dot{x}_1, \dot{x}_2, \theta_1, \theta_2, \dot{\theta}_1, \dot{\theta}_2)$, with $x_i \in [-1, 1]$ being the position and $\dot{x}_i \in [-10, 10]$ the velocity of the $i$-th cart, and with $\theta_i \in [0, 2\pi]$ being the angle and $\dot{\theta}_i \in [-5, 5]$ being the angular velocity of the $i$-th pole. The two dimensional control vector is discretized to the four actions $a \in \{(-2, -2), (-2, +2), (+2, -2), (+2, +2)\}$ and held constant for $\Delta t = 0.1$sec. The dynamics of the system and physical parameters we used to instantiate the problem are specified in Appendix B.

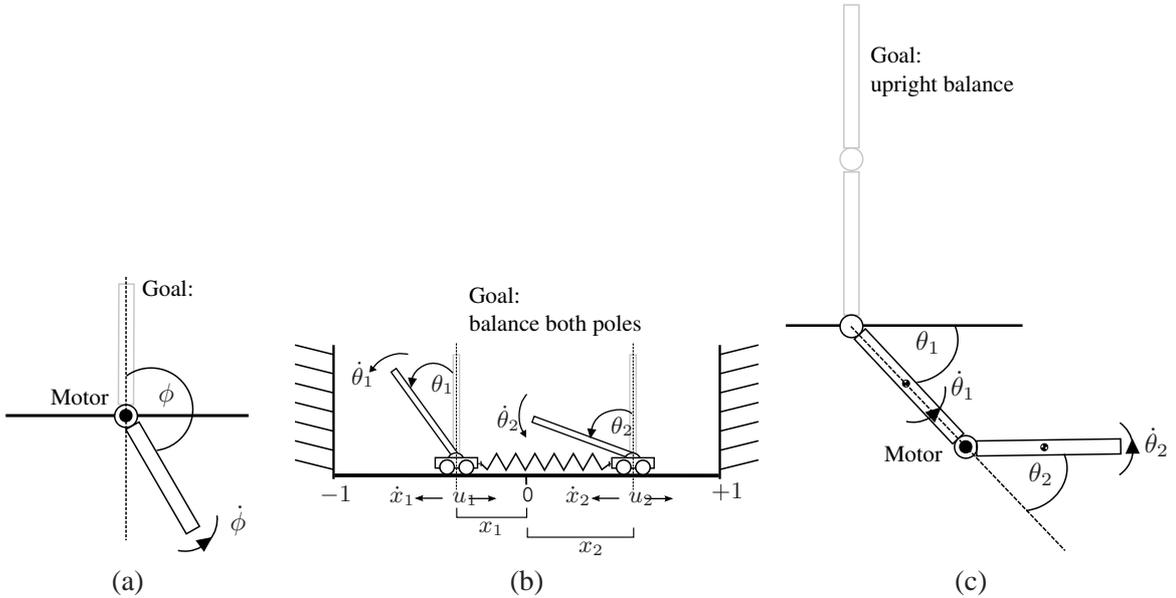

Figure 5: From left to right: the inverted pendulum task, the double inverted pendulum linked with a spring task, and the acrobot handstand task.

To formulate this task as a reinforcement learning problem of the form given in Eq. (5), we used the following settings: the set of initial states is the singleton $X_0 = \{(0, 0.5, \pi, \pi, 0, 0, 0, 0)\}$, the reward is defined as $\varrho(x(t), u(t)) := [(1 + \cos\theta_1(t)) + (1 + \cos\theta_2(t))]/4$, the discount factor is set to $\gamma = 0.999$, and each policy is evaluated for $H = 250$ steps. In order to achieve a high reward in this domain, the policy has to both balance the poles and not collide with one of the walls. Low rewards usually occur because of such collisions.

**Acrobot** Our third domain is the acrobot from [46]: a two-link robot that resembles a gymnast swinging up above a high bar (see Figure 5c). The acrobot freely swings around the first joint (the hands grasping the bar) and can exert force only at the second joint (bending the hips). The acrobot is an underactuated system; the task we consider here is the inverted "handstand" position, which is hard to solve using reinforcement learning methods. The state space is 4-dimensional, $x \equiv (\theta_1, \theta_2, \dot\theta_1, \dot\theta_2)$, with $\theta_1$ and $\theta_2$ being the angle of the upper and lower link, and $\dot\theta_1$ and $\dot\theta_2$ their angular velocity, respectively. The continuous control is discretized to $\{-1, +1\}$ and held constant for $\Delta t = 0.2$sec. To facilitate staying stable in the inverted handstand position (a highly unstable equilibrium), we also include a third non-primitive "balance" action, which chooses control values derived from an LQR controller obtained from linearizing the system dynamics about the handstand position. Note that this balance action produces meaningful outputs $\in [-1, +1]$ only very close to the unstable equilibrium and thus cannot be used to bring the acrobot from the initial state to the goal region. The dynamics of the system and physical parameters we used to instantiate the problem are specified in Appendix C.

To formulate this task as a reinforcement learning problem of the form given in Eq. (5), we used the following settings: the set of initial states is the singleton $X_0 = \{(-\pi/2, 0, 0, 0)\}$, the reward is defined by the height of the end of the second link (the feet) as $\varrho(x(t), u(t)) = 2 + \cos(\theta_1(t) - \pi/2) + \cos(\theta_1(t) + \theta_2(t) - \pi/2)$ (+100 if handstand), the discount factor is set to $\gamma = 1$, and each policy is evaluated for $H = 500$ steps[6].

**HIV drug treatment** Our last problem domain is taken from a real-world application in medical control [1]. The aim is to optimize the treatment of a patient infected by HIV over a period of a few years using what is known as structured treatment interruption (STI). The treatment of the patient consists of choosing a combination of two drugs, yielding 4 possible choices (including the case where no drug at all is taken), to administer every 5 days. Administering the correct cocktail with the correct timing can hinder the spread of HIV infected cells and will eventually bring the patient into a healthy state (a locally stable equilibrium); however, the drugs also have side effects on the patient's health and thus their use should be kept to a minimum. Finding an optimal treatment strategy is considered a challenging optimal control problem with highly nonlinear transition dynamics [16]. The system is represented by a six-dimensional state vector $x \equiv (T_1, T_2, T_1^*, T_2^*, V, E)$, where $T_1 \geq 0$ and $T_2 \geq 0$ is the count of healthy type-1 and type-2 cells, $T_1^* \geq 0$ and $T_2^* \geq 0$ is the count of infected type-1 and type-2 cells, $V \geq 0$ is the number of free virus copies, and $E \geq 0$ the number of immune response cells. The two-dimensional control vector $u \equiv (\varepsilon_1, \varepsilon_2)$ consists of the dosage of two drugs which is discretized to the four actions $\{(0.3, 07), (0.7, 0), (0, 0.3), (0, 0)\}$ and held constant for $\Delta t = 5$ days. The dynamics of the system and physical parameters we used to instantiate the problem are specified in Appendix D.

---

[6] Note that since $\gamma = 1$, the criterion that we optimize for this problem is the (time-independent) finite sum of rewards. Changing the criterion in this way does not make any technical difference for any of the direct policy search techniques.

To formulate this task as a reinforcement learning problem of the form given in Eq. (5), we used the following settings: the set of initial states is the singleton $X_0 = \{(163573, 5, 11945, 46, 63919, 24)\}$, which corresponds to the unhealthy locally stable equilibrium (i.e., a high number of HIV-infected cells). The reward is composed from the number of infected and uninfected cells plus an additional term reflecting the cost for using a drug: $\varrho(x(t), u(t)) = -0.1V(t) + 10000E(t) - 20000\varepsilon_1(t) - 20000\varepsilon_2(t)$. The discount factor is set to $\gamma = 0.98$, and each policy is evaluated for $H = 300$ steps. Note that in this domain, unlike in all the previous ones, the reward is not upper-bounded (and thus the optimistic tree development strategy from Eq. (12) cannot be applied). Moreover, the values of the state variables can vary over a large range, from 0 up to the order of $10^6$; to counter any unwanted scaling effects in our learning methods we transformed the state variables by taking their logarithm.

## 5.2 Contestant methods

We compare OLT policies against both pure DPS approaches and pure LT approaches. From the point of view of DPS, we consider three alternative policy representations (neural networks, adaptive radial basis functions and optimized look-ahead trees) and two alternative optimizers (cross-entropy and GPO). Our pure LT approaches are composed of the uniform tree development strategy, the optimistic development strategy proposed in [23] and two greedy tree development strategies.

**DPS Representations.** We consider the following two common policy representations:

- *Neural network.* This representation is probably the most widely used in the direct policy search literature [4]. The policy is represented by a fully connected feed-forward neural network, which has one hidden layer and one output layer with one neuron per possible action. Given the current state, the neural network computes one activation score per action and the policy returns an action with maximal score. Hidden nodes have *tanh* activations and their number is a hyper-parameter that enables to control the complexity of the policy; output nodes have a linear activation function.

- *Adaptive radial basis functions.* In this representation proposed in [6], a policy is encoded through a set of basis functions that are attached to particular actions. Given the current state, the policy works by searching for the nearest basis function ("nearest" being measured under the Mahanalobis distance metric) and by returning the action attached to this basis function. Each single basis function is parameterized by the location of its center, the length-scales along each dimension and the recommended action. The number of basis functions used to encode the policy is a complexity hyper-parameter that has to be tuned.

Table 1 summarizes the characteristics of the three policy representations that we consider in this paper. Each representation has a hyper-parameter that enables to control the policy complexity. Note that optimized look-ahead trees are unique in the sense that the number of their parameters – hence the complexity of the associated direct policy search problem – do not depend on the value of this hyper-parameter.

**DPS Optimizers.** Cross-entropy (CE) [44] is a versatile global optimization technique that is widely used in direct policy search [4]. In particular, this method was used in previous work [32], under the alternative name of *estimation of distribution algorithm*. The chief advantage of cross-entropy (and the main reason for it being highly popular) is that it is very easy to implement and, because it is population-based, can deal with a potentially very large number of sample locations (i.e., CE operates only on small batches of sample locations the size of which is constant throughout, whereas GPO has to consider all current sample locations to determine the next one). The algorithm works by fitting a distribution to the best currently found solutions and by using this distribution to sample new candidate solutions. In our case, we use a simple variant of cross-entropy that relies on a multi-variate Gaussian distribution with *diagonal* covariance matrix. This distribution is first initialized to cover the whole search space: we start with a Gaussian with zero mean and a diagonal covariance matrix, the entries of which are equal to the the square of the half-length of an interval centered at zero and containing the corresponding coordinate of the search space. The algorithm then draws a number $N_{CE}$ of observations from this distribution, where each observation is a parameter vector and represents a possible policy, and evaluates the $N_{CE}$ resulting policies and sorts them according to their performance. It then picks a number $M_{CE}$ of the top best performing policies, and uses them to update the generating distribution: the old mean is replaced by the sample mean and each diagonal entry of the covariance matrix is replaced by the per-coordinate variance of the $M_{CE}$ best parameter vectors. These two steps are iterated until a stopping condition, in our case a predefined maximum number of iterations, is reached.

| Name | Policy representation | Hyper-parameter | Number of parameters |
|---|---|---|---|
| Neural network (NN) | Feed forward neural network with $\tanh$ activation and one hidden layer | Number of hidden neurons ($nHidden$) | $(n_x + 1) \times nHidden + (nHidden + 1) \times K$ |
| Radial basis functions (RBF) | Adaptive radial basis functions with Mahalanobis distance as in [6] | Number of basis functions ($nBF$) | $2 \times n_x \times nBF$ |
| Optimized look-ahead trees (OLT) | Look-ahead tree (algo 2) with optimized node scoring heuristic $e(\cdot; \theta)$ as in Eq. (13) | Budget of node expansions ($h_{\max}$) | $3 \times n_x$ |

Table 1: Summary of the policy representations compared in this paper. $n_x$ denotes the dimensionality of the state space and $K$ is the number of actions.

**LT strategies.** Classical look-ahead tree policies rely on fixed generic node scoring functions. In order to demonstrate the significance of learning the node selection rule in a problem-driven way, we compare our approach to look-ahead tree policies using $e^{uniform}$ (uniform tree development, Eq. 11), $e^{optimistic}$ (the method proposed by [23], Eq. 12) and two forms of greedy tree development:

$$e^{greedy-1}(n_{i,h}) := \varrho_{t+h} \qquad e^{greedy-2}(n_{i,h}) := \gamma^h \varrho_{t+h}. \tag{26}$$

## 5.3 Performance comparison

We start by comparing the performance we obtain with the different approaches discussed previously. Note that, contrarily to OLT, NN and RBF typically involve solving challenging global optimization problems with hundreds or thousands of parameters. While GPO is quite efficient for problems that have a reasonable number of parameters, this approach requires sophisticated approximations to scale to higher-dimensional problems (e.g., see [38]). In this paper we use a naive textbook implementation of GPO that is able to solve OLT optimization problems, but that suffers from scaling problems when the number of samples increases, which happens to be problematic in high-dimensional problems. Therefore, we use CE to learn NN and RBF based policies. To make the comparison fair, we also use CE to optimize the parameters of OLT policies in this first part of our empirical study. A comparison between CE and GPO for learning OLT policies is provided in Section 5.4.

**Experimental protocol.** We use the same test procedure for all policies and the performance is measured as the discounted sum of rewards obtained when executing the policy for $H$ steps (see Section 5.1), starting from the initial state[7] $x_0$.

The parameters of CE were tuned by hand so as to give in each case a good result with a reasonable amount of computation. The result of this tuning is given in Table 2. Note that, since OLT involves lower-dimensional optimization problems than the two other representations, the required optimization budget ($N_{CE} \times nIter$) is lower on all problem for this representation. For the RBF representation, the CE has to optimize both continuous parameters (RBF centers and lengthscales) and discrete parameters (action assignments). To optimize these discrete parameters, we chose a multinomial distribution with Dirichlet prior where the initial count of each action was set to 10.

---
[7]Recall that for each of our four benchmark domains the set $X_0$ only contains one single initial state $x_0$. Yet as we will see below, robustness with respect to perturbations of the initial state is one of the strengths of look-ahead tree policies, so that optimizing only over one initial state is justified.

|       |          | Inverted pendulum | | | Double inverted pendulum | | | Acrobot handstand | | | HIV drug treatment | | |
|-------|----------|:---:|:---:|:---:|:---:|:---:|:---:|:---:|:---:|:---:|:---:|:---:|:---:|
|       |          | $N_{CE}$ | $M_{CE}$ | nIter | $N_{CE}$ | $M_{CE}$ | nIter | $N_{CE}$ | $M_{CE}$ | nIter | $N_{CE}$ | $M_{CE}$ | nIter |
| NN    |          | 400 | 20 | 100 | 400 | 20 | 200 | 1000 | 20 | 50  | 400 | 20 | 200 |
| RBF   |          | 200 | 20 | 100 | 500 | 10 | 200 | 500  | 10 | 200 | 500 | 20 | 200 |
| OLT   |          | 100 | 10 | 25  | 100 | 10 | 50  | 100  | 10 | 50  | 100 | 10 | 50  |

Table 2: Cross-entropy parameters for the four problems and the three policy representations.

|     |             | Inverted pendulum | Double inverted pendulum | Acrobot handstand | HIV drug treatment |
|-----|-------------|:---:|:---:|:---:|:---:|
| NN  | nHidden     | 50  | 50  | 20  | 15  |
|     | Performance | 89.9 | 134.9 | 3.89e4 | 0.88e9 |
| RBF | nBF         | 30  | 6   | 30  | 15  |
|     | Performance | 73.4 | 122.5 | **4.09e4** | 1.39e9 |
| OLT | Budget      | 31  | 1365 | 40 | 85 |
|     | Performance | **93.2** | **145.2** | 4.07e4 | **4.22e9** |

Table 3: Quality of learned policies with neural networks, radial basis functions and optimized look-ahead trees. For each problem, the best performance is shown in bold. For each method and each problem we also display the value of the tuned hyper-parameter.

The hyper-parameters of the three policy representations were tuned by grid-search. For both the number of hidden nodes in the NN representation and the number of basis functions in the RBF representation, we tested the following values: $\{1, 2, 3, 4, 5, 7, 10, 12, 15, 20, 30, 40, 50\}$. As in [23, 32], we tested budget values for look-ahead tree policies that correspond to the number of nodes of fully developed trees of varying depth $d = 1, \ldots, 8$. These budgets values are as follows: $h_{\max} \in \{1, 1 + K, 1 + K + K^2, 1 + K + K^2 + K^3, \ldots\}$.

**Performance comparison.** Table 3 reports the best scores obtained by the tree kinds of policies as well as the values of the tuned hyper-parameters. We observe that OLT works significantly better than the other DPS methods on three out of the four benchmark domains. On the HIV domain, OLT enables us to obtain slightly better results than the state of the art [16] (4.22e9 against 4.16e9), whereas both of the two other representations only manage to reach policies with very moderate scores ($\approx 1e9$). RBF slightly outperforms OLT on the acrobot domain. However, as we will see later, this result holds thanks to a very careful tuning of the number of radial basis functions, whereas OLT works well for a wide range of budget values.

**Relevance of learning the node scoring function.** We have seen that the OLT policy representation enables to reach policies outperforming those obtained with neural networks and radial basis functions. One could wonder whether this result could be obtained by using look-ahead trees without learning. We therefore performed a series of experiments that, for various budget values, compare the performance of look-ahead tree policies, with and without learning. Note that in these experiments, the budget is set before optimization, hence we optimize one OLT policy per tested budget value.

The results of our comparison between OLT policies and traditional LT policies is given in Figure 6. Best scores are obtained with OLT policies on three domains out of the four (inverted pendulum, acrobot and HIV). Furthermore, thanks to learning, significantly lower budgets are required to reach a given level of performance, again on three domains out of the four (double inverted pendulum, acrobot and HIV). The most impressive results are obtained on the HIV domain, for which an OLT policy with a budget of 2 node expansions already performs better than all other LT policies with a budget up to $h_{\max} = 87381$ node expansions (which corresponds to fully developed trees of depth 8). An OLT with such a small budget consists in first expanding the node corresponding to the current state and then expanding one of its successor nodes. Our results show that carefully selecting this successor state is much more efficient to solve the HIV problem than developing large trees in a generic way. In the same spirit, on the acrobot problem, an OLT policy with only 8 node expansions performs slightly better than a uniformly developed tree with a budget 9841 nodes (corresponding to a tree of depth of 7).

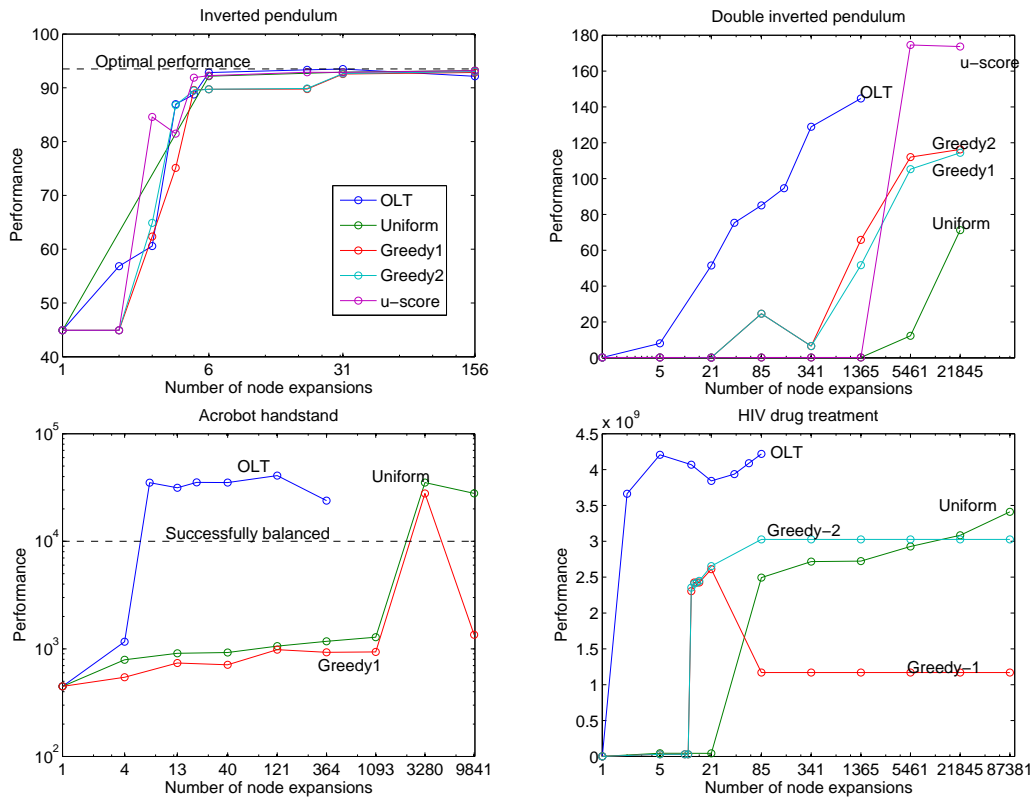

Figure 6: Performance of optimized look-ahead tree policies (OLT) vs. baseline look-ahead tree policies for various budget values. Since $\gamma = 1$ in the acrobot domain, the greedy-2 policy degenerates to greedy-1 and the $u$-score based policy degenerates to the uniform policy. Since the reward is not upper-bounded in the HIV domain, the $u$-score is not defined for this problem. As explained in the text, because the time required to make a single decision increases linearly in the number of node expansions and because we have to evaluate many more policies to produce a single OLT curve than we have to produce a LT curve, we did not evaluate OLT on the same large budget values as LT.

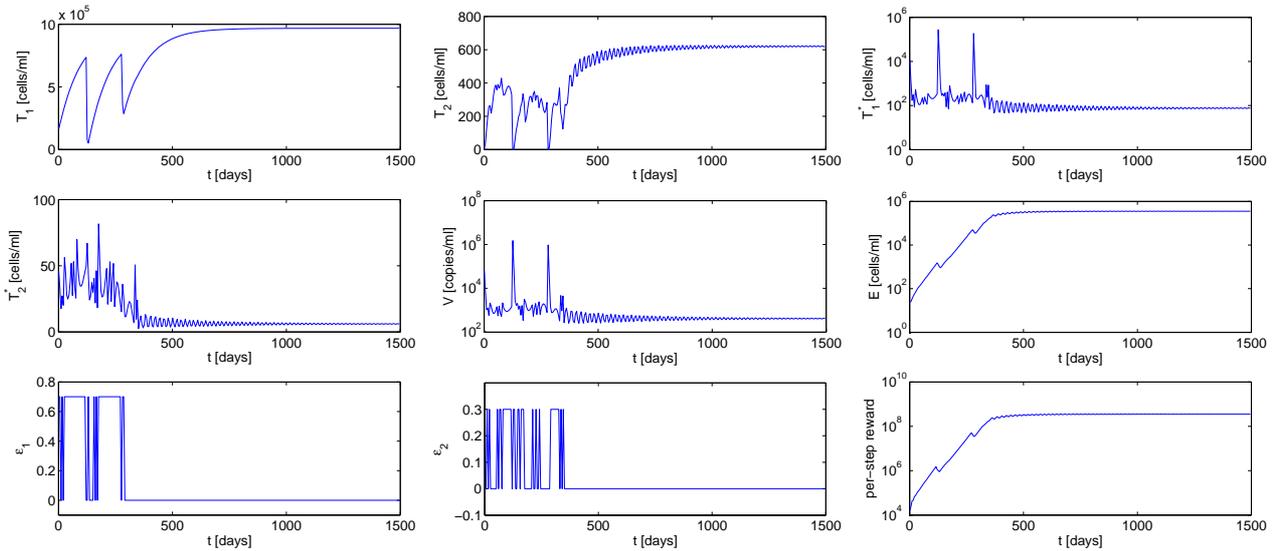

Figure 7: Trajectory of the HIV system when controlled by an OLT policy with a budget of 85 node expansions. The first six panels show the development of the six system states, the following two panels show the dosage of the RTI and PI drugs applied, and the last panel shows the reward obtained in each step.

On the double inverted pendulum problem, OLTs are ultimately outperformed by look-ahead tree policies using $u$-score. However, we observe that our approach is much better able to deal with a constrained computational budget. For example, using only 341 node expansions we obtain a reasonably well performing policy, that outperforms all other generic tree development strategies even when they have very large budgets.

On the inverted pendulum problem, we see that all LT policies achieve a near-optimal performance (which we can compute for this domain, e.g. see [25]) already with a budget of 5. It thus seems that there is little interest in learning a specific node scoring function for this problem.

**Illustration of the HIV policy.** In order to allow a direct comparison between the performance of our method on the HIV domain with the performance given in the earlier related work [16, 7, 6], we plot in Figure 7 the trajectory that we obtain on the HIV system. These results show that our policy applies RTI and PI drugs in a way which is very similar to what other state-of-the-art policies do (which, however, are obtained in a fundamentally different manner, e.g., by fitted Q-value iteration using millions of sample transitions).

### 5.4 Sample efficiency

We have seen that the OLT representation enables to reach high-performance policies, which often outperform alternative DPS representations. Besides performance, another aspect of major importance in DPS is the sample efficiency of the learning process, i.e. how fast good policies can be obtained. We study sample efficiency by looking at the performance of our different methods in function of two metrics: the number of policy evaluations performed and the number of transitions simulated.

**Experimental protocol.** The most common solution to measure sample efficiency in a DPS scheme is to look at the number of policy evaluations required to reach a certain level of performance. This measure corresponds to the number of different parameter values that have been tried by the optimization algorithm. Here, we consider two different optimization algorithms: cross-entropy and GPO. Cross-entropy is tuned as previously and GPO is instantiated as follows: as kernel we chose, as it is standard for GP regression, the squared exponential given in Eq. (18), where the hyperparameters are found for each batch of data via marginal likelihood optimization [42]. In this optimization, the best setting of the hyper-parameters found in the previous iteration is used as the mean of the hyperprior in the next iteration. To generate an initial batch of training data, we generated 10 samples (100 samples in the double inverted pendulum domain) via Latin hypercube sampling; these initial samples are taken into account when we compare sample efficiency. To find at each iteration of GPO the best next sample location, we optimize the EI acquisition function from Eq. (25) with $\zeta = 0.01$ using DIRECT.

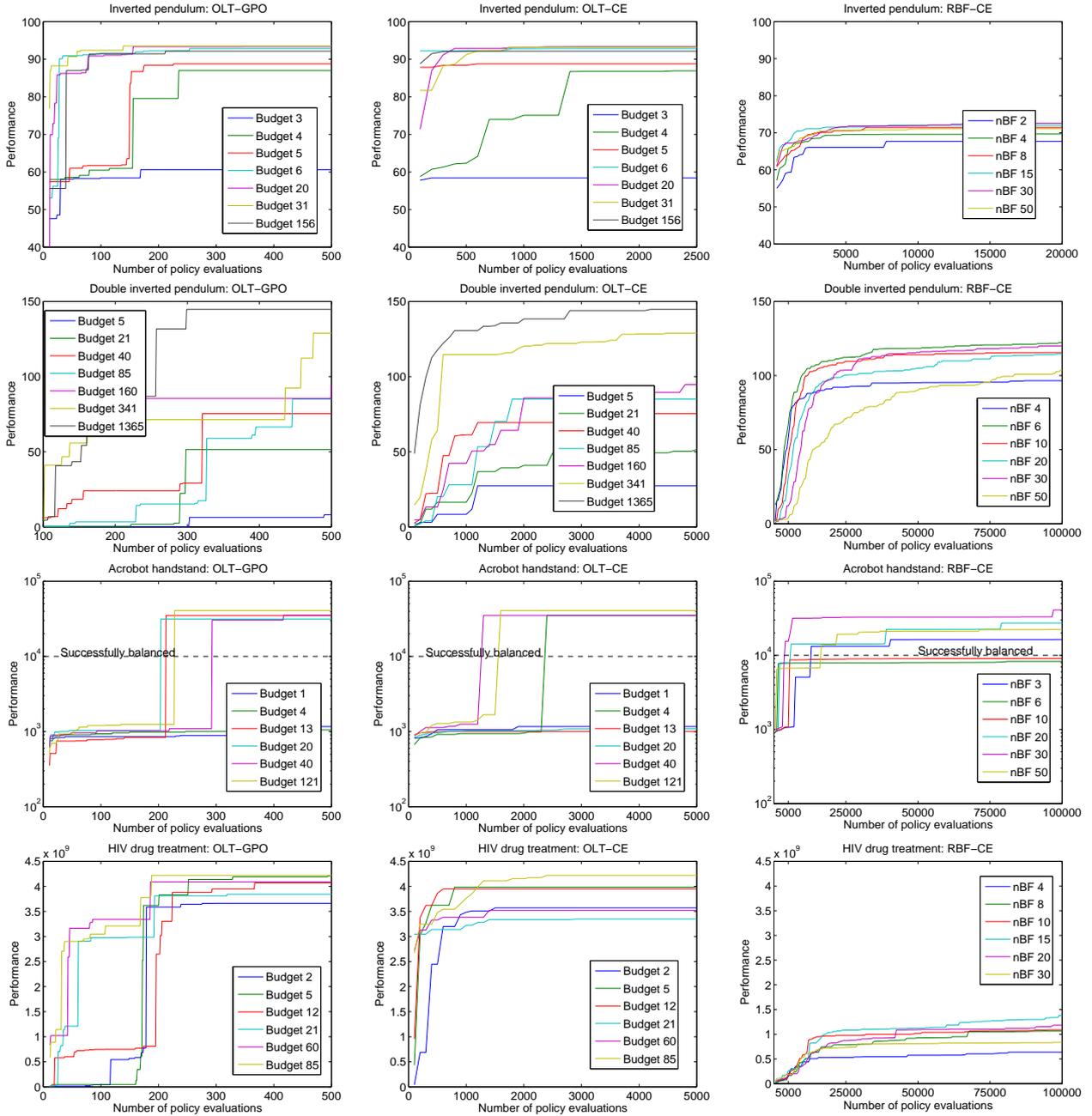

Figure 8: Performance vs. number of policy evaluations. From left to right: OLT with GPO, OLT with CE and RBF with CE. From top to bottom: inverted pendulum, double inverted pendulum, acrobot handstand and HIV drug treatment.

We are interested in two questions: how do GPO compare against CE when using the OLT representation? and how does the choice of representation impact the sample efficiency? To answer these questions, we selected the following three setups: OLT with GPO, OLT with CE and RBF with CE, and plotted the corresponding learning curves for different values of the $h_{\max}$ and $nBF$ hyper-parameters. These plots are given in Figure 8.

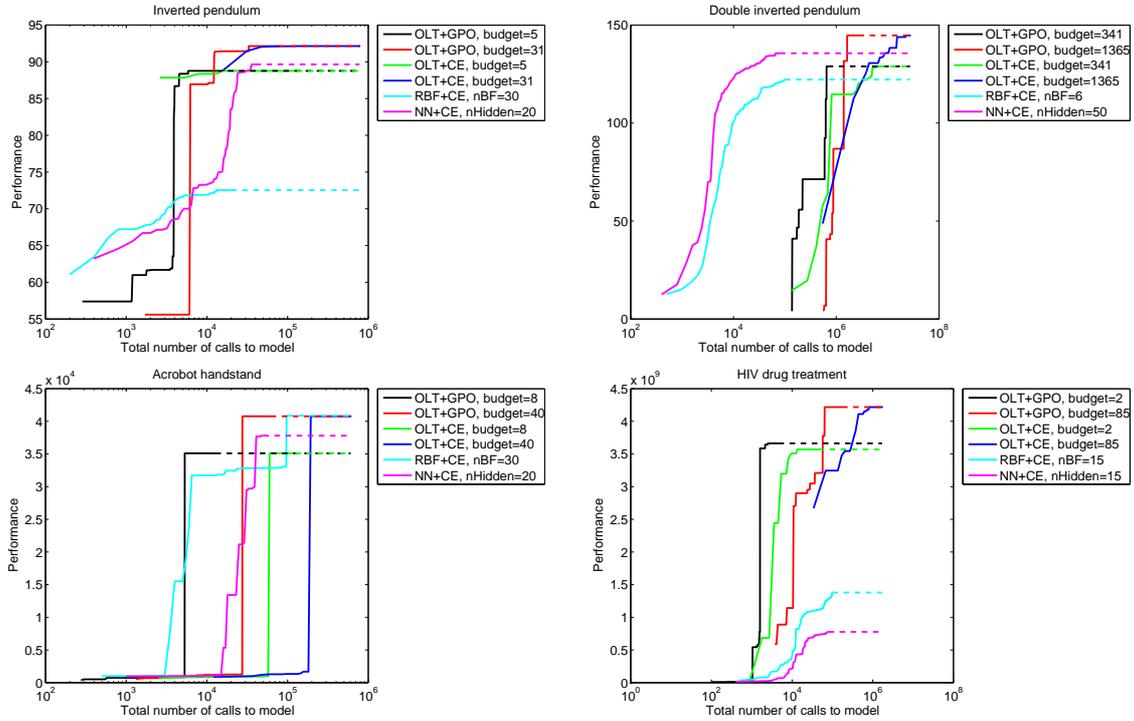

Figure 9: Performance vs. number of simulated transitions.

**Impact of the optimizer.** We observe that GPO requires about one order of magnitude fewer policy evaluations than CE on all four problems (note that the GPO curves stop at 500 policy evaluations, whereas we display CE results until 5000 evaluations). We observe no notable difference between the performance of the final policies obtained with GPO and CE, and the overall shapes of the learning curves are similar. Note that CE is much easier to implement than GPO. Therefore, in the light of these results, we suggest using CE when sample complexity is not a big problem and selecting GPO in the case where the number of policy evaluations is a strong applicative constraint. Using GPO requires a bit more work, but is profitable since it enables to roughly perform learning one order of magnitude faster.

**Impact of the representation.** Our results show that optimizing look-ahead trees is substantially more sample efficient than optimizing the parameters in a basis function representation, this difference being 5K versus 50K-100K policy evaluations when using the cross-entropy method. The same kinds of differences are observed when comparing against the NN representation. One obvious explanation for this is that optimizing the node scoring function involves fewer parameters, which results in the search space having lower dimensionality. As mentioned previously, a key advantage of OLT is that the complexity of the policy (i.e. the computational budget $h_{\max}$) can be arbitrarily scaled without increasing the number of parameters to be optimized. In contrast, for both the neural network and adaptive basis function representation higher complexity can only be achieved at the expense of a larger number of policy parameters, which generally involves harder optimization problems.

**Performance vs. number of simulated transitions.** An important characteristic of OLT policies is that they use some online computational resources to take their decisions. Specifically, in order to take a single decision, a look-ahead tree policy requires simulating $Kh_{\max}$ transitions using a generative model of the problem, since computing the policy involves expanding $h_{\max}$ nodes and since expanding a node invokes the model once per possible action. Hence, it is also important to study the performances of the algorithm with respect to the *number of simulated transitions*. This new sample complexity measure is computed in the following way. Note that one policy evaluation requires making a trajectory of $H$ steps (assuming we have a unique training initial state) and that the simulator is called once per step of this trajectory. The number of simulated transitions per policy evaluation is thus $H$ for NN and RBF policies. If we add the additional simulation cost of OLT policies, the number of simulated transitions per policy evaluation becomes $H \times (1 + Kh_{\max})$.

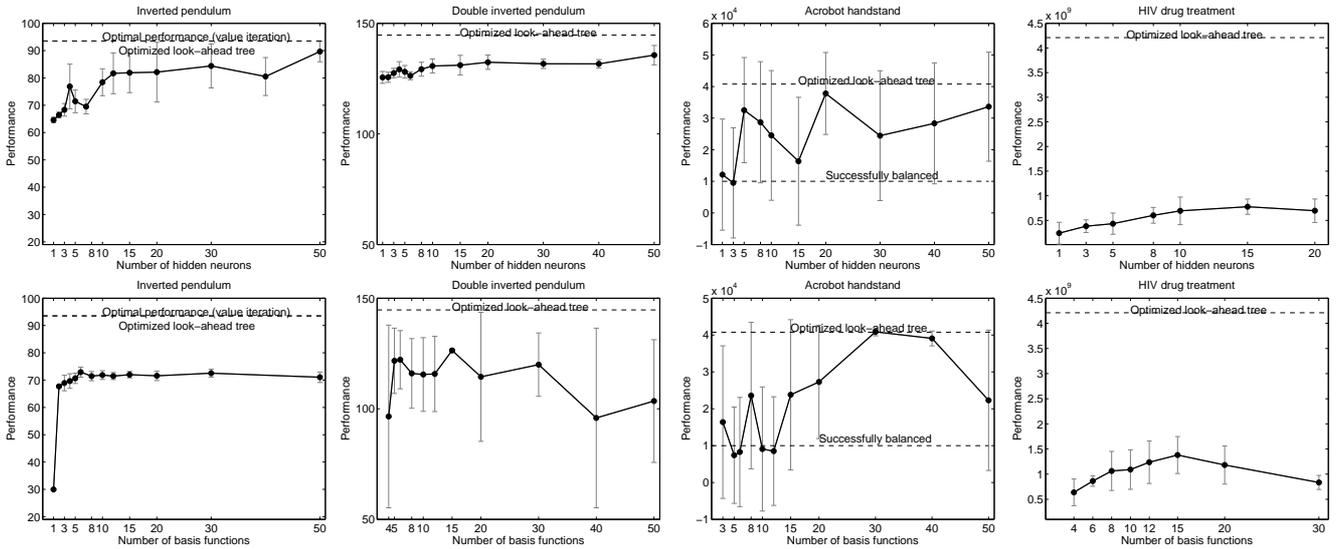

Figure 10: Performance vs. hyper-parameter value. Top: Performance of NN in function of the number of hidden neurons. Bottom: Performance of RBF in function of the number of basis functions.

We compare our three policy representations with the tuned hyper-parameters given in Table 3. Since OLT policies with high budget values are strongly disadvantaged w.r.t. the number of simulated transitions, we consider an additional *low-buget* OLT setting, where the budget values were chosen by hand to be as small as possible while still producing reasonably good policies asymptotically. Figure 9 reports the learning curves obtained by CE for the NN and RBF settings and by both GPO and CE for the two OLT settings.

We observe slightly different behaviors depending on the problem. In the two first problems, the best policies are obtained with NN and RBF for small number of simulated transitions, and after a given threshold, OLT policies become better. This was expected since evaluating a single OLT policy can already require simulating thousands of transitions. In the acrobot and HIV problems, OLT policies outperform the other ones on almost the whole x-axis range, which was more unexpected. This proves that the larger number of transitions required to take decisions is partly compensated by the smaller number of parameters to optimize.

On all four problems, we observe that (i) low-budget OLT policies outperform tuned OLT policies when the number of simulated transitions is small, (ii) policies optimized with GPO always outperform their counterparts optimized with CE, and (iii) the asymptotically best (or almost best in the case of acrobot) policies are based on optimized look-ahead trees.

### 5.5 Robustness with respect to hyper-parameters

In addition to performance and learning efficiency, another important property of DPS approaches from a practical perspective is that their hyper-parameters should be easy to tune. Indeed, tuning these parameters is part of the whole policy learning process, hence the more trial and error is required, the longer it takes to obtain high-performance policies. The behavior of OLT policies in function of the $h_{\max}$ hyper-parameter was shown in Figure 6. We report the performance of the other two kinds of policies in function of their respective hyper-parameters in Figure 10. Since CE is a stochastic optimization algorithm, we performed these experiments ten times for each setting and report the empirical means and standard deviations.

Depending on the problem, NN and RBF are more or less sensitive to the setting of their complexity parameter. For both methods, there are some problems for which tuning the complexity hyper-parameter is rather easy, and others for which it is much harder. We observe that on the acrobot benchmark, which is the only one where NN and RBF are able to produce policies competitive with OLT policies, both the performance of NN and RBF policies have large variances: performance varies strongly both for increasing the complexity (number of nodes/basis functions) or across different runs in the same setting. To understand this large variance, it should be noted that acrobot is different from the previous domains in that there is no partial solution: a policy is either able to swing up and successfully balance in the handstand position indefinitely (the reward is $> 10^4$) or not at all (the reward is of order $10^3$ at most). Successful balance is only possible if the system approaches the tiny region in the state space from where LQR can take over.

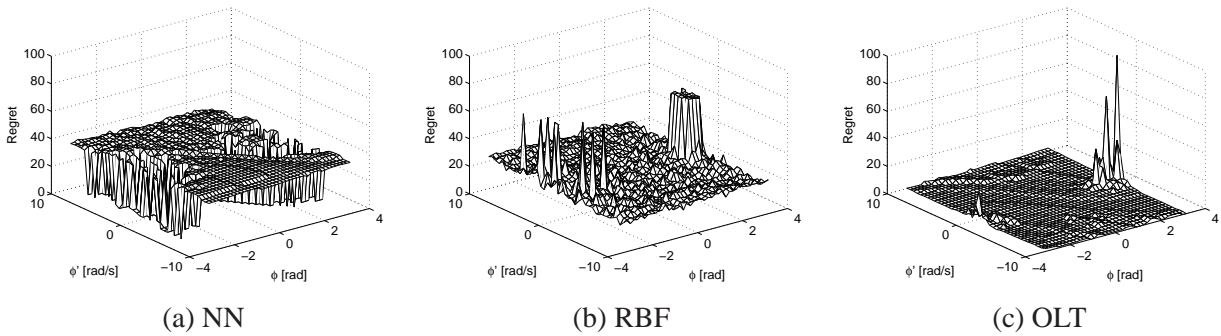

(a) NN  (b) RBF  (c) OLT

Figure 11: Regret for NN, RBF and OLT policies on the inverted pendulum domain when the testing initial state differs from the training initial state.

Although we observe in some cases that well-performing policies can be found using a fairly compact representation (e.g., in the double inverted pendulum domain, already one hidden neuron can give a good performance), our results indicate that, seen across all domains, finding the best possible complexity parameter for NN and RBF requires a larger amount of trial and error experimentation. At the opposite, as shown in Figure 6, OLTs are rather easy to tune, since, in general, increasing the budget increases the quality of the policy or only slightly degrades it. In all our experiments, a default value corresponding to a fully developed tree of depth 3 or 4 yielded good performing policies.

### 5.6 Robustness with respect to initial states.

In DPS, policies are optimized w.r.t. a particular distribution over initial states. An important issue is that this distribution may not perfectly match future unexpected usage of the learned policy. In general, a preference is given to policies that are robust w.r.t. such mismatches. To compare the robustness of the NN, RBF and OLT policies and study how well they are able to generalize when starting from initial states which differ from those used during learning, we performed experiments by evaluating them on perturbed or completely different initial states on the inverted pendulum problem and the HIV drug treatment problem.

**Stability on inverted pendulum.** To analyze the stability of the various policies on this domain, we started by learning NN, RBF and OLT policies with the single initial state $x_0 = (-\pi, 0)$ and then evaluated these policies across the whole domain by systematically varying the initial state. Remember that we are able to compute a truly optimal policy on the inverted pendulum domain. We use this latter policy to report the *regret* for each initial state, i.e. the difference between the optimal performance and performance gotten from the learned policy. The results of these experiments are given in Figure 11. It can be seen that an OLT policy (budget 20) performs well and incurs close to zero regret across large parts of the state space, with an exception being the region around $(\pi, 0)$. When comparing with NN and RBF policies, we see that both representations are less able to generalize and can incur comparatively large regrets even for small perturbations.

**Stability on HIV drug treatment.** Figure 12 illustrates the robustness of an OLT policy (budget 85) in the HIV domain. This time we are no longer able to exhaustively sample the state space; instead we perturb each state variable in turn by multiplying each variable of $x_0$ independently by a factor ranging from 0.1 to 10. The figure then shows that the performance stays largely stable even for large perturbations[8]. In particular, in all cases the performance of the OLT policy remains significantly higher than the best results obtained by NN and RBF ($> 3e9$ vs. $\approx 1e9$).

---

[8]Note that in doing this we did not pay attention to the physical meaning of the variables; thus by increasing the value of the sixth state variable it appears as if the policy begins to perform even better. However, since this state variable denotes the count of healthy cells in the blood of a patient, increasing its value by an order of magnitude alters the whole problem.

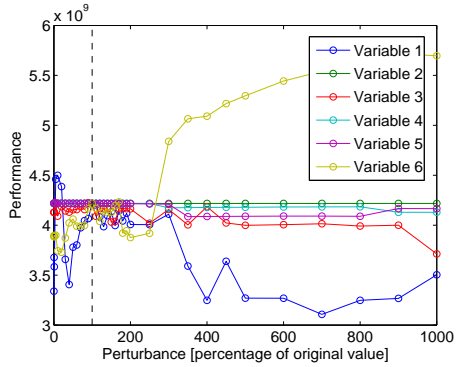

Figure 12: Robustness of an OLT policy for the HIV drug treatment domain. The plot examines how the performance of the policy (with budget 85) decreases when the initial state for which the policy was optimized is perturbed: each curve corresponds to the perturbation of one state variable independently. The vertical line at $x = 100$ denotes the unperturbed value.

## 6 Related work

The approach proposed in this paper lies at the intersection of two families of solutions for sequential-decision making. We first overview *direct policy search* in Section 6.1 and then discuss relevant work on *look-ahead tree search* in Section 6.2. Section 6.3 positions our approach with respect to Model Predictive Control techniques. Finally, Section 6.4 suggests that optimized look-ahead tree policies belong to a larger emergent class of techniques: *parameterized algorithms for decision-making*.

### 6.1 Direct policy search

Direct policy search is a widely used class of solutions that comes from reinforcement learning. For a general overview over the field of reinforcement learning, refer to one of the books of [48, 2, 41, 4]. Over the years, many DPS techniques have been proposed and giving credit to every one of them would be hardly possible. Individual techniques differentiate themselves by their policy parametrization and the optimization method used for identifying parameters leading to a high-performing policy. The main distinction is whether the policy search is *gradient-free* or *gradient-based*.

Gradient-based DPS, which very often also goes by the name of policy-gradient method, follows an iterative optimization scheme and uses the gradient of the value function to adapt and change the policy parameters such that performance increases. This requires the value function to be a smooth function of the policy parameters, which is typically achieved by considering compatible stochastic policies. Often however, the gradient cannot be computed analytically since the value function itself is not available in closed form. Instead, the gradient has to be estimated using, e.g., finite difference approximation, Monte-Carlo rollouts or value function approximation which leads to the actor-critic methods. One of the best known examples is probably the natural policy gradient method described in [40]; another more recent example is the work in [14]. While impressive results have been obtained in particular for learning controllers in robotics, policy-gradient methods do have some weaknesses; the most notable ones being a high sensitivity to the initial value of the policy parameters (the starting point of the iteration), abundance of local minima, and difficulties when the return is noisy or stochastic. Note that in this paper we do not deal with gradient-based policy search.

Gradient-free DPS on the other hand finds the best policy parameters via derivative-free global optimization. Its main strength is simplicity and generality; the latter meaning that, because it is derivative-free, the representation of a policy by any algorithm with adjustable parameters is admissible, and not only those for which the value function will be differentiable. Since gradient-free DPS techniques perform a global search in the parameters space, they do not suffer from local minima or from the problem of having to guess a good initial solution; these techniques were also shown to cope well with stochastic returns and hidden states. For certain domains in reinforcement learning such as Tetris, the best performing policies known today have been obtained by gradient-free DPS (see [49] and follow-up work). The weakness of this approach is that conceptually it is less sample-efficient than policy-gradient methods and thus will require a substantially higher number of policy evaluations. This comes as a consequence of having to solve a global optimization problem over a potentially high-dimensional search space; however, by carefully designing the policy parametrization and choosing powerful optimizers sample efficiency is often improved.

Early examples of DPS came under the guise of evolutionary approaches to reinforcement learning, or neuroevolution, and consisted of neural networks as policy representation, with the weights making up the policy parameters, and variants of genetic algorithms acting as global optimizer. Examples can be found in [37] and [18], where later work also considered optimization of the network structure [47], or using recurrent neural networks to better cope with hidden states [19]. A recent comparison of these methods can also be found in [51] and [26].

As an alternative to genetic algorithms, more recent work started to explore the use of the cross-entropy method [44], or variants such as the covariance matrix adaptation evolution strategy CMA-ES [20] as global optimizer for policy search. Examples include [22], where the policy was represented by a simple linearly parametrized function, [26], where the policy was represented by a neural network, or [6], where the policy was represented by adaptive radial basis functions. These latter two were also used in this paper as baseline methods during the experimental evaluation of our look-ahead trees. Another example for a policy parametrization is to use domain-specific building blocks, such as motor primitives, as it was done in [27] and [29] to optimize the gait of the AIBO quadrupedal robot. A different kind of policy representation is used in [50] to learn a policy for the game of Ms. Pac-Man; here the policy is represented by a list of domain-specific parameterized rules.

Section 4 described a third option for the global optimization part: Gaussian process optimization [35, 3]. GPO can achieve very good sample efficiency and was previously considered for policy search in [29] and [17]. In the end however, the choice of which global optimizer one uses will always be secondary; what is more important is how the policy is represented and to what extent this representation facilitates the optimization process by shaping the "fitness landscape".

## 6.2 Look-ahead tree search

The algorithm studied in this paper can also be related to the larger field of tree-based planning and search. One of the most seminal works in this field is the $A^*$ algorithm [21] which uses a best-first search to find the shortest path from a source state/configuration to a goal state/configuration. The conceptual difference between $A^*$ and related methods on the one side and what we are presenting here on the other side is that our method implements an *online* and *finite computational budget* mechanism: a search tree is grown at each decision-making step, and only a finite number of node expansions are allowed during tree development (the larger the computational budget, the better the decision). In $A^*$ the search tree needs to be grown until a goal state is reached. Our method is thus capable of online planning and producing closed-loop policies, whereas $A^*$ is not. More specifically, the algorithm studied in this paper can also be interpreted as a method for *learning* an exploration strategy in a tree. In $A^*$, the function used for evaluating the nodes is the sum of two terms: the length of the so far shortest path from the source to the current node and an optimistic estimate of the shortest path from this node to the goal (a so-called "admissible" heuristic). Several authors have sought to learn good admissible heuristics for the $A^*$ algorithm. For example, we can mention the LRTA$^*$ algorithm [28] which is a variant of A$^*$ and which learns over multiple trials an optimal admissible heuristic. More recent work for learning strategies to efficiently explore graphs have focused on the use of supervised regression techniques using various approximation structures to solve this problem (e.g., linear regression, neural networks, $k$-nearest neighbors); for example, see [30, 34, 52, 21]. The objective of which node to best explore next also arises in the context of game-playing; here we can mention, e.g., Monte-Carlo tree search [12]. In particular progressive strategies which widen up the actions (nodes considered for expansion) such as in [45, 12, 9] and which can also be used for continuous action spaces [43] could be a promising enhancement for the policy search technique presented herein.

## 6.3 Model predictive control

Model Predictive Control techniques have originally been introduced as ways to stabilize large-scale systems with constraints around equilibrium points (or around a reference trajectory) [36, 10, 15]. They exploit an explicitly formulated model of the problem and solve in a receding horizon manner a series of finite time open-loop deterministic optimal control problems. In such they are very much related to look-ahead tree techniques. Actually, a MPC technique that searches for the first action of the optimal sequence of actions over the finite optimisation horizon through (clever) exploration of a tree is a look-ahead tree technique. However, most of the control techniques labelled as MPC techniques do not use tree-exploration for identifying (near-) optimal sequences of actions. They rather assume strong regularity assumptions on the system dynamics and the reward function (e.g., linearity) – that we do not make here – which are exploited to reformulate the search for an optimal open-loop sequence of action as a standard mathematical programming problem (e.g., convex optimisation problem, mixed integer programming problem). This reformulation of the problem as a mathematical programming problem often leads to techniques that are able to scale well to very large state-action spaces, provided of course that the regularity assumptions hold.

### 6.4 Parameterized algorithms for decision-making

Our work relies on a rather simple and generic research methodology: for a given targeted class of problems, (i) identify an algorithm skeleton that is believed to provide promising solutions to problem instances, (ii) parameterize this algorithm and (iii) optimize the parameters in a problem-driven way, through DPS, i.e., through direct optimization of the algorithm performance. This methodology can be applied to a wide range of problem kinds and has already inspired several authors. The system proposed in [5] places an optimisation layer on top of an approximate value iteration algorithm to optimize the location of its basis functions. In [13] (Section 5.3), the authors consider multi-stage stochastic programming techniques and optimize the scenario trees using Monte-Carlo methods. Closer to our work, it is proposed in [11] to parameterize a tree-search technique for decision-making: *upper confidence trees*. In this work, the parameters enable to control the simulation policy used to estimate long-term returns within the upper confidence tree algorithm.

Parameterized algorithms have also been shown to be relevant to solve various kinds of exploration / exploitation dilemma in a problem-driven way. References [33] and [31] propose to learn exploration / exploitation strategies for multi-armed bandit problems either by using the same kind of parameterizations (a simple linear function) and the same kind of optimizers (derivative-free global optimizers) as ours or by searching in a space of formulas. Reference [8] extends this idea to the exploration / exploitation dilemma that occurs in single-trajectory reinforcement learning. Here, the parameters are no more real-valued vectors, but rather small formulas that depend on the internal variables of the policy. Given a target class of Markov decision problems, this method is able to learn high-performance policies that outperform state-of-the-art generic reinforcement learning algorithms.

We believe that with the recent progress in derivative-free global optimization (with algorithms such as cross-entropy, CMA-ES and GPO) and the continuously growing available computing power, the class of such parameterized algorithms optimized in a problem-driven way has high chances to grow in the near future, since they offer a systematic way to improve upon generic solutions, by exploiting problem-dependent characteristics through learning.

## 7 Conclusion

This paper has introduced *optimized look-ahead tree policies, a novel technique* that bridges the gap between two major families of solutions for solving sequential decision-making problems: direct policy search (DPS) and look-ahead tree (LT) policies. Our approach uses a new way of representing a policy by using parameterized look-ahead trees. In this representation, the parameters of the policy over which optimization takes place encode the node scoring function by which the tree is grown (until a predefined computational budget is exhausted) every time an action is required from the system. This approach manages to combine the best of both DPS and LT approaches. From the point of view of DPS, optimized look-ahead tree policies provide a highly-generic way of representing policies that alleviates the need to choose a complex function approximator. From the point of view of LT, DPS is the key that enables to significantly reduce the online computational requirement, through a principled offline parameter-learning procedure.

We have shown through an extensive experimental study that our approach has several qualities: it (1) produces high-performance policies that often outperform both pure LT policies and pure DPS policies, (2) requires at the offline training stage a substantially smaller number of policy evaluations than other DPS techniques, thanks to a small parameter space whose size does not depend on the policy complexity, (3) is easy to tune since it avoids having to choose a particular function approximator and (4) results in policies that are quite robust with respect to perturbations of the initial states. Compared to traditional LT policies, we have shown that learning a node scoring function can result in substantial savings in terms of required online computational resources.

This paper has focused on a particular kind of sequential decision-making problems: finite actions, deterministic transitions, known transition and reward function (i.e., a generative black-box model which allows us to simulate arbitrary transitions), and an "informative" reward. However, we believe that except for the last item, all the other items on this list can be addressed in future work. For example, large number of actions means that we can no longer exhaustively generate all successor states when predicting one step ahead; instead, one would have to use sampling techniques or progressive widening techniques which is also an active research topic in tree-based search [45, 12, 9, 43]. Deterministic transitions can be relaxed to weakly stochastic transitions (weakly meaning that there is only a small number of possible successor states) as was done in [7] to extend the optimistic strategy from [23] to sparse stochastic systems. And finally, in application scenarios where a generative model is not available from the beginning, one could also try to integrate model-learning over time into the policy search framework (e.g., see [25]).

Table 4: Physical parameters of the inverted pendulum domain

| Symbol | Value | Meaning |
| --- | --- | --- |
| $g$ | 9.81 $[m/s^2]$ | gravitation |
| $m$ | 1 $[kg]$ | mass of link |
| $l$ | 1 $[m]$ | length of link |
| $\mu$ | 0.05 | coefficient of friction |

# Acknowledgments

Tobias Jung acknowledges financial support from a research fellowship of ULg. Damien Ernst acknowledges the financial support of the Belgian National Fund of Scientific Research (FNRS) of which he is a Research Associate. This paper presents research results of the PASCAL2 European FP7 Network of Excellence, and of the BIOMAGNET Network funded by the Interuniversity Attraction Poles Programme initiated by the Belgian State, Science Policy Office.

# A  Dynamic model of the inverted pendulum

Refer to the schematic representation of the inverted pendulum given in Figure 5a. The state variables are the angle measured from the vertical axis, $\phi(t)$ [rad], and the angular velocity $\dot\phi(t)$ [rad/s]. The control variable is the torque $u(t)$ [Nm] applied, which is restricted to the interval $[-5, 5]$. The motion of the pendulum is described by the differential equation:

$$\ddot\phi(t) = \frac{1}{ml^2}\Big(-\mu\dot\phi(t) + mgl\sin\phi(t) + u(t)\Big). \tag{27}$$

The angular velocity is restricted via saturation to the interval $\dot\phi \in [-10, 10]$. The values and meaning of the physical parameters are given in Table 4.

The solution to the continuous-time dynamic equation is obained by writing Eq. (27) as a first-order system and using a Runge-Kutta solver with $h = 40$ intermediate steps. The time step of the simulation is $\Delta t = 0.2$ sec, during which the applied control is kept constant. The 2-dimensional state vector is $x(t) \equiv \big(\phi(t), \dot\phi(t)\big)$, the scalar control variable is $u(t)$. Since our algorithm requires a finite set of possible actions, we discretized the continuous control space into 5 discrete action choices $a \in \{-5, -2.5, 0, 2.5, 5\}$.

# B  Dynamic model of the double inverted pendulum

Refer to the schematic representation of the double inverted pendulum on spring-linked carts given in Figure 5b. The state variables are, for each cart $i = 1, 2$, the angle of displacement measured from the vertical axis, $\theta_i(t)$ [rad], the angular velocity $\dot\theta_i(t)$ [rad/s], the position of the cart $x_i(t)$ [m] measured from the origin (note $x_1(t) < x_2(t)$), and its velocity $\dot x_i(t)$ [m/s]. The vector-valued control is the force $u_i(t)$ [Nm] applied to each cart, which is restricted to the interval $[-2, 2]$. The system as a whole is described by the system of differential equations

Table 5: Physical parameters of the double inverted pendulum domain

| Symbol | Value | Meaning |
|---|---|---|
| $g$ | $9.81\ [m/s^2]$ | gravitation |
| $L$ | $1.0\ [m]$ | half-length of the track |
| $l$ | $0.5\ [m]$ | half-length of a pole |
| $m_c$ | $1.0\ [kg]$ | mass of a cart |
| $m_p$ | $0.1\ [kg]$ | mass of a pole |
| $\mu_c$ | $0.0005$ | coefficient of friction of a cart |
| $\mu_p$ | $0.000002$ | coefficient of friction of a pole |
| $K$ | $2.0$ | coefficient $K$ of the spring |
| $l_s$ | $0.5\ [m]$ | relaxed length of the spring |
| $l_{smin}$ | $0.1\ [m]$ | minimum length of the spring before deformation |
| $l_{smax}$ | $1.5\ [m]$ | maximum length of the spring before deformation |

(see online appendix of [23]):

$$\ddot{\theta}_i(t) = \left[b_i^2(t)a_i^{12}(t) - a^{22}b_i^1(t)\right]/\left[a_i^{12}(t)a_i^{21}(t) - a^{11}a^{22}\right], \qquad i = 1, 2 \tag{28}$$

$$\ddot{x}_i(t) = \left[b_i^1(t) - a^{11}\ddot{\theta}_i(t)\right]/a_i^{12}(t), \qquad i = 1, 2, \tag{29}$$

where

$$\begin{aligned}
a_i^{12}(t) &:= -\cos\theta_i(t) \\
a_i^{21}(t) &:= lm_p\cos\theta_i(t) \\
b_i^1(t) &:= g\sin\theta_i(t) - \mu_p\dot{\theta}_i(t)/(lm_p) \\
b_i^2(t) &:= lm_p\dot{\theta}_i^2(t)\sin\theta_i(t) - f_i(t) + \mu_c\,\text{sign}(\dot{x}_i(t)) \\
f_i(t) &:= u_i(t) + K(l_s - |x_2(t) - x_1(t)|) \\
a^{11} &:= 4l/3 \\
a^{22} &:= -(m_c + m_p).
\end{aligned}$$

The angular velocity is restricted via saturation to the interval $\dot{\phi}_i \in [-10, 10]$, the velocity of the cart to $\dot{x}_i \in [-5, 5]$. The system knows two terminal conditions which lead to a stopping of the process: collision between one of the carts and a wall, and collision between the carts. More specifically, these conditions are implemented as follows: the temporal evolution of the system is halted if at any time $t$ at least one of the following is true:

- $|x_i(t)| > L$ (first cart collides with left wall or second cart collides with right wall)
- $x_2(t) \leq x_1(t)$ (first cart has passed the second cart)
- $|x_2(t) - x_1(t)| \notin [l_{smin}, l_{smax}]$ (outside minimal and maximal length of spring before deformation)

The values and meaning of the physical parameters are given in Table 5.

The solution to the continuous-time dynamic equations is obtained by writing Eqs. (28)-(29) as a first order system and using a Runge-Kutta solver with $h = 10$ intermediate steps. The time step of the simulation is $\Delta t = 0.1$ sec, during which the applied control is kept constant. The 8-dimensional state vector is $x(t) \equiv \bigl(x_1(t), x_2(t), \dot{x}_1(t), \dot{x}_2(t), \theta_1(t), \theta_2(t), \dot{\theta}_1(t), \dot{\theta}_2(t)\bigr)$, the control vector is $u(t) \equiv \bigl(u_1(t), u_2(t)\bigr)$. Since our algorithm requires a finite set of possible actions, we discretized the control space into 4 discrete action choices $a \in \{(-2, -2), (-2, +2), (+2, -2), (+2, +2)\}$.

Table 6: Physical parameters of the acrobot domain

| Symbol | Value | Meaning |
|---|---|---|
| $g$ | $9.8\,[m/s^2]$ | gravitation |
| $m_i$ | $1\,[kg]$ | mass of link $i$ |
| $l_i$ | $1\,[m]$ | length of link $i$ |
| $l_{ci}$ | $0.5\,[m]$ | length to center of mass of link $i$ |
| $I_i$ | $1\,[kg \cdot m^2]$ | moment of inertia of link $i$ |

## C  Dynamic model of the acrobot

Refer to the schematic representation of the acrobot domain given in Figure 5c. The state variables are the angle of the first link measured from the horizontal axis, $\theta_1(t)$ [rad], the angular velocity $\dot{\theta}_1(t)$ [rad/s], the angle between the second link and the first link $\theta_2(t)$ [rad], and its angular velocity $\dot{\theta}_2(t)$ [rad/s]. The control variable is the torque $\tau(t)$ [Nm] applied at the second joint. The dynamic model of the acrobot system is [46]:

$$\ddot{\theta}_1(t) = -\frac{1}{d_1(t)}\bigl(d_2(t)\ddot{\theta}_2(t) + \phi_1(t)\bigr) \tag{30}$$

$$\ddot{\theta}_2(t) = \frac{1}{m_2 l_{c2}^2 + I_2 - \frac{d_2(t)^2}{d_1(t)}}\left(\tau(t) + \frac{d_2(t)}{d_1(t)}\phi_1(t) - m_2 l_1 l_{c2}\dot{\theta}_1(t)^2 \sin\theta_2(t) - \phi_2(t)\right) \tag{31}$$

where

$$d_1(t) := m_1 l_{c1}^2 + m_2\bigl(l_1^2 + l_{c2}^2 + 2l_1 l_{c2}\cos\theta_2(t)\bigr) + I_1 + I_2$$
$$d_2(t) := m_2\bigl(l_{c2}^2 + l_1 l_{c2}\cos\theta_2(t)\bigr) + I_2$$
$$\phi_1(t) := -m_2 l_1 l_{c2}\dot{\theta}_2(t)^2 \sin\theta_2(t) - 2m_2 l_1 l_{c2}\dot{\theta}_2(t)\dot{\theta}_1(t)\sin\theta_2(t) + \bigl(m_1 l_{c1} + m_2 l_1\bigr)g\cos\theta_1(t) + \phi_2(t)$$
$$\phi_2(t) := m_2 l_{c2} g \cos\bigl(\theta_1(t) + \theta_2(t)\bigr).$$

The angular velocities are restricted via saturation to the interval $\theta_1 \in [-4\pi, 4\pi]$, and $\theta_2 \in [-9\pi, 9\pi]$. The values and meaning of the physical parameters are given in Table 6; we used the same parameters as in [48].

The solution to the continuous-time dynamic equations in Eqs. (30)-(31) is obained using a Runge-Kutta solver with $h = 20$ intermediate steps. The time step of the simulation is $\Delta t = 0.2$ sec, during which the applied control is kept constant. The 4-dimensional state vector is $x(t) \equiv \bigl(\theta_1(t), \theta_2(t), \dot{\theta}_1(t), \dot{\theta}_2(t)\bigr)$, the scalar control variable is $\tau(t)$.

The motor was allowed to produce torques $\tau$ in the range $[-1, 1]$. Since our algorithm requires a finite set of possible actions, we discretized the continuous control space. Here we use three actions: the first two correspond to a bang-bang control and take on the extreme values $-1$ and $+1$. However, a bang-bang control alone does not allow us to keep the acrobot in the inverted handstand position, which is an unstable equilibrium. As a third action, we therefore introduce a more complex balance-action, which is derived via LQR. First, we linearize the acrobot's equation of motion about the unstable equilibrium $(-\pi/2, 0, 0, 0)$, yielding:

$$\dot{x}(t) = Ax(t) + Bu(t),$$

where, after plugging in the physical parameters of Table 6,

$$A = \begin{bmatrix} 0 & 0 & 1 & 0 \\ 0 & 0 & 0 & 1 \\ 6.21 & -0.95 & 0 & 0 \\ -4.78 & 5.25 & 0 & 0 \end{bmatrix}, \quad B = \begin{bmatrix} 0 \\ 0 \\ -0.68 \\ 1.75 \end{bmatrix}, \quad x(t) = \begin{bmatrix} \theta_1(t) - \pi/2 \\ \theta_2(t) \\ \dot{\theta}_1(t) \\ \dot{\theta}_2(t) \end{bmatrix} \quad u(t) = \tau(t).$$

Using MATLAB, an LQR controller was then computed for the cost matrices $Q = I_{4\times 4}$ and $R = 1$, yielding the state feedback law

$$u(t) = -Kx(t), \tag{32}$$

with constant gain matrix $K = [-189.28, -47.46, -89.38, -29.19]$. The values resulting from Eq. (32) were truncated to stay inside the valid range $[-1, 1]$. Note that the LQR controller works as intended and produces meaningful results only when the state is already in a close neighborhood of the handstand state; in particular, it is incapable of swinging up and balancing the acrobot on its own from the initial state $(0, 0, 0, 0)$.

## D  Dynamic model of the HIV drug treatment domain

The HIV infection dynamics are described by a six-dimensional nonlinear system with the state vector $x(t) \equiv \big(T_1(t), T_2(t), T_1^*(t), T_2^*(t), V(t), E(t)\big)$, where

1. $T_1(t) \geq 0$ ($T_1^*(t) \geq 0$) is the number of non-infected (infected) $CD4^+$ T-lymphocytes (in cells/ml),

2. $T_2(t) \geq 0$ ($T_2^*(t) \geq 0$) is the number of non-infected (infected) macrophages (in cells/ml),

3. $V(t) \geq 0$ is the number of free HI virus particles (in copies/ml), and

4. $E(t) \geq 0$ is the number of cytotoxic T-lymphocytes (in cells/ml).

The dynamics is described by the following system of first-order differential equations (see [1]):

$$\dot{T}_1(t) = \lambda_1 - d_1 T_1(t) - (1 - \varepsilon_1(t))k_1 V(t) T_1(t) \tag{33}$$

$$\dot{T}_2(t) = \lambda_2 - d_2 T_2(t) - (1 - f\varepsilon_1(t))k_2 V(t) T_2(t) \tag{34}$$

$$\dot{T}_1^*(t) = (1 - \varepsilon_1(t))k_1 V(t) T_1(t) - \delta T_1^*(t) - m_1 E(t) T_1^*(t) \tag{35}$$

$$\dot{T}_2^*(t) = (1 - f\varepsilon_1(t))k_2 V(t) T_2(t) - \delta T_2^*(t) - m_2 E(t) T_2^*(t) \tag{36}$$

$$\dot{V}(t) = (1 - \varepsilon_2(t))N_T \delta (T_1^*(t) + T_2^*(t)) - cV(t) - \\ \big[(1 - \varepsilon_1(t))\rho_1 k_1 T_1(t) + (1 - f\varepsilon_1(t))\rho_2 k_2 T_2(t)\big]V(t) \tag{37}$$

$$\dot{E}(t) = \lambda_E + \frac{b_E(T_1^*(t) + T_2^*(t))}{T_1^*(t) + T_2^*(t) + K_b}E(t) - \frac{d_E(T_1^*(t) + T_2^*(t))}{T_1^*(t) + T_2^*(t) + K_d}E(t) - \delta_E E(t) \tag{38}$$

The vector-valued control variable is $u(t) = (\varepsilon_1(t), \varepsilon_2(t))$, where $\varepsilon_1$ and $\varepsilon_2$ corresponds to the dosage of the reverse transcriptase inhibitor drug (RTI) and the protease inhibitor drug (PI), respectively. In STI, drugs are either fully administered (they are "on") or not at all (they are "off"). A fully administered RTI drug corresponds to the value $\varepsilon_1 = 0.7$, while a fully administered PI drug corresponds to the value $\varepsilon_2 = 0.3$. This leads to a discrete actions space with four possible choices $a \in \{(0.7, 0.3), (0.7, 0), (0, 0.3), (0, 0)\}$. Because it is not clinically feasible to change the treatment daily, the state is measured and the drugs are switched on or off once every five days. Therefore, the system is controlled in discrete time with a sampling period of $\Delta t = 5$ days (during which the chosen controls are kept constant).

As shown in [1], in the absence of treatment (i.e. $\varepsilon_1 = \varepsilon_2 \equiv 0$), the system in Eqs. (33)-(37) exhibits three physical equilibrium points:

1. an unstable equiliberium point $(T_1, T_2, T_1^*, T_2^*, V, E) = (10^6, 3198, 0, 0, 0, 10)$ which represents an uninfected state;

2. a "healthy" locally stable equilibrium point $(T_1, T_2, T_1^*, T_2^*, V, E) = (967839, 621, 76, 6, 415, 353108)$ which corresponds to a small viral load, a high $CD4^+$ T-lymphocytes count and a high HIV-specific cytotoxic T-cells count;

3. a "non-healthy" locally stable equilibrium point $(T_1, T_2, T_1^*, T_2^*, V, E) = (163573, 5, 11945, 46, 63919, 24)$ for which T-cells are depleted and the viral load is very high.

Numerical simulations show that the basin of attraction of the healthy steady-state is relatively small in comparison with the one of the non-healthy steady-state. Furthermore, perturbation of the uninfected steady-state by adding as little as one single particle of virus per *ml* of blood plasma leads to asymptotical convergence towards the non-healthy steady-state.

The solution to the continuous-time dynamic equations in Eqs. (33)-(37) is obtained by using a Runge-Kutta solver with $h = 500$ intermediate steps. The values and meaning of the constants in the model is the same as in [1],[16]: $\lambda_1 = 10,000$, $d_1 = 0.01$, $k_1 = 8 \cdot 10^{-7}$, $\lambda_2 = 31.98$, $d_2 = 0.01$, $f = 0.34$, $k_2 = 1 \cdot 10^{-4}$, $\delta = 0.7$, $m_1 = 1 \cdot 10^{-5}$, $m_2 = 1 \cdot 10^{-5}$, $N_T = 100$, $c = 13$, $\varrho_1 = 1$, $\varrho_2 = 1$, $\lambda_E = 1$, $b_E = 0.3$, $K_b = 100$, $d_E = 0.25$, $K_d = 500$, $\delta_E = 0.1$.

# References

[1] B. Adams, H. Banks, H.-D. Kwon, and H. Tran. Dynamic multidrug therapies for HIV: optimal and STI control approaches. *Mathematical Biosciences and Engineering*, 1:223–241, 2004.

[2] D. Bertsekas. *Dynamic programming and Optimal Control, Vol. II.* Athena Scientific, 2007.

[3] E. Brochu, V. M. Cora, and N. de Freitas. A tutorial on bayesian optimization of expensive cost functions, with application to active user modeling and hierarchical reinforcement learning. *CoRR*, abs/1012.2599, 2010.

[4] L. Busoniu, R. Babuska, B. De Schutter, and D. Ernst. *Reinforcement Learning and Dynamic Programming using Function Approximators*. Taylor & Francis CRC Press, 2010.

[5] L. Busoniu, D. Ernst, R. Babuska, and B. De Schutter. Fuzzy partition optimization for approximate fuzzy q-iteration. In *Proceedings of the 17th IFAC World Congress (IFAC-08)*, 2008.

[6] L. Busoniu, D. Ernst, R. Babuska, and B. De Schutter. Cross-entropy optimization of control policies with adaptive basis functions. *IEEE Transactions on Systems, Man, and Cybernetics-Part B: Cybernetics*, 41(1):196–209, 2011.

[7] L. Busoniu, R. Munos, B. De Schutter, and R. Babuska. Optimistic planning for sparsely stochastic systems. In *Proc. of IEEE International Symposium on Adaptive Dynamic Programming and Reinforcement Learning (ADPRL-11)*, pages 48–55, 2011.

[8] M. Castronovo, F. Maes, R. Fonteneau, and D. Ernst. Learning exploration/exploitation strategies for single trajectory reinforcement learning. In *Proc. of 10th European Workshop on Reinforcement Learning*, Edinburgh, Scotland, June 2012.

[9] G. Chaslot, M. Winands, J. Uiterwijk, H. van den Herik, and B. Bouzy. Progressive strategies for monte-carlo tree search. In *Proc. of the 10th Joint Conf. on Information Sciences*, 2007.

[10] T. Coen, J. Anthonis, and J. De Baerdemaeker. Cruise control using model predictive control with constraints. *Comput. Electron. Agric.*, 63(2):227–236, October 2008.

[11] A. Couetoux, H. Doghmen, and O. Teytaud. Improving the exploration in upper confidence trees. In *Proc. of the 6th Learning and Intelligent OptimizatioN Conf.*, 2012.

[12] R. Coulom. Efficient selectivity and backup operators in monte-carlo tree search. In *Proc. of the 5th Int. Conf. on Computers and Games*, 2006.

[13] B. Defourny, D. Ernst, and L. Wehenkel. Multistage stochastic programming: a scenario tree based approach to planning under uncertainty. In Eduardo F. Morales L. Enrique Sucar and Jesse Hoes, editors, *Decision Theory Models for Applications in Artificial Intelligence: Concepts and Solutions*. ISBN: 978-1609601652, 2012.

[14] M. P. Deisenroth and C. E. Rasmussen. Pilco: A model-based and data-efficient approach to policy search. In *Proc. of the International Conference on Machine Learning (ICML 2011)*, 2011.

[15] D. Ernst, M. Glavic, F. Capitanescu, and L. Wehenkel. Reinforcement learning versus model predictive control: a comparison on a power system problem. *IEEE Transactions on Systems, Man, and Cybernetics - Part B: Cybernetics*, 2009.

[16] D. Ernst, G. Stan, J. Goncalves, and L. Wehenkel. Clinical data based optimal STI strategies for HIV: a reinforcement learning approach. In *Proc. of the 45th IEEE Conference on Decision and Control*, 2006.

[17] M. Frean and P. Boyle. Using Gaussian processes to optimize expensive functions. In W. Wobcke and M. Zhang, editors, *AI 2008: Advances in Artificial Intelligence, vol 5360 LNCS*, pages 258–267. Springer: Berlin, 2008.

[18] F. Gomez and R. Mikkulainen. Active guidance for a finless rocket using neuroevolution. In *Proc. of the Genetic and Evolutionary Computation Conference (GECCO 2003)*, pages 2084–2095, 2003.

[19] F. Gomez, J. Schmidhuber, and R. Mikkulainen. Accelerated neuroevolution through cooperatively coevolved synapses. *Journal of Machine Learning Research*, 9:937–965, 2008.

[20] N. Hansen. The CMA evolution strategy: a comparing review. In J. A. Lozano, P. Larranaga, I. Inza, and E. Bengoetxea, editors, *Towards a new evolutionary computation. Advances on estimation of distribution algorithms*, pages 75–102. Springer, 2006.

[21] P. Hart, N. Nilsson, and B. Raphael. A formal basis for the heuristic determination of minimum cost paths. *IEEE Transactions on Systems Science and Cybernetics*, 4(2):100–107, 1968.

[22] V. Heidrich-Meisner and C. Igel. Variable metric reinforcement learning methods applied to the noisy mountain car problem. In *Proc. of European Workshop on Reinforcement Learning (EWRL 2008)*, pages 136–150, 2008.

[23] J-F. Hren and R. Munos. Optimistic planning of deterministic systems. In *Proc. of European Workshop on Reinforcement Learning*, pages 151–164. Springer, 2008.

[24] D. R. Jones. A taxonomy of global optimization methods based on response surfaces. *Journal of Global Optimization*, 21:345–383, 2001.

[25] T. Jung and P. Stone. Gaussian processes for sample efficient reinforcement learning with rmax-like exploration. In *Proc. of ECML*, 2010.

[26] S. Kalyanakrishnan and P. Stone. Characterizing reinforcement learning methods through parameterized learning problems. *Machine Learning*, 82(3):205–247, 2011.

[27] N. Kohl and P. Stone. Machine learning for fast quadrupedal locomotion. In *Proc. of the 19th National Conference on Artificial Intelligence (AAAI 2004)*, pages 611–616, 2004.

[28] R. E. Korf. Real-time heuristic search. *Artificial Intelligence*, 42(2-3):189–211, 1990.

[29] D. Lizotte, T. Wang, M. Bowling, and D. Schuurmans. Automatic gait optimization with gaussian process regression. In *Proc. of IJCAI*, 2007.

[30] F. Maes. *Learning in Markov Decision Processes for Structured Prediction*. PhD thesis, Pierre and Marie Curie University, Computer Science Laboratory of Paris 6 (LIP6), October 2009.

[31] F. Maes, L. Wehenkel, and D. Ernst. Automatic discovery of ranking formulas for playing with multi-armed bandits. In *Proc. of the 9th European Workshop on Reinforcement Learning (EWRL 2011)*, 2011.

[32] F. Maes, L. Wehenkel, and D. Ernst. Optimized look-ahead tree policies. In *Proceedings of the 9th European Workshop on Reinforcement Learning (EWRL 2011)*, 2011.

[33] F. Maes, L. Wehenkel, and D. Ernst. Learning to play K-armed bandit problems. In *Proc. of International Conference on Agents and Artificial Intelligence*, Vilamoura, Algarve, Portugal, February 2012.

[34] S. Minton. *Machine Learning Methods for Planning*. Morgan Kaufmann Publishers Inc., San Francisco, CA, USA, 1994.

[35] J. Mockus. *Bayesian Approach to Global Optimization*. Kluwer Academic Publishers, 1989.

[36] M. Morari and J.H. Lee. Model predictive control: Past, present and future. *Computers & Chemical Engineering*, 23(4):667–682, 1999.

[37] D. E. Moriarty, A.C. Schultz, and J.J. Grefenstette. Evolutionary algorithms for reinforcement learning. *Journal of Artificial Intelligence Research*, 11:241–276, 1999.

[38] M. Osborne, R. Garnett, and S. J. Roberts. Gaussian processes for global optimization. In *Proc. of 3rd International Conference on Learning and Intelligent Optimization (LION 3)*, pages 1–15, 2009.

[39] C. D. Perttunen, D. R. Jones, and B. E. Stuckman. Lipschitzian optimization without the lipschitz constant. *Journal of Optimization Theory and Application*, 79(1):157–181, 1993.

[40] J. Peters and S. Schaal. Natural actor-critic. *Neurocomputing*, 71(7-9):1180–1190, 2008.

[41] W. Powell. *Approximate Dynamic Programming*. Wiley, 2007.

[42] C. E. Rasmussen and C. K. I. Williams. *Gaussian Processes for Machine Learning*. MIT Press, 2006.

[43] P. Rolet, M. Sebag, and O. Teytaud. Optimal active learning through billiards and upper confidence trees in continuous domains. In *Proc. of ECML*, 2009.

[44] R. Y. Rubinstein and D. P. Kroese. *The Cross Entropy Method: A Unified Approach to Combinatorial Optimization, Monte-Carlo Simulation, and Machine Learning*. Springer, 2004.

[45] N. Sokolovska, O. Teytaud, and M. Milone. Q-learning with double progressive widening: application to robotics. In *Proc. of ICONIP*, pages 103–112, 2011.

[46] M. Spong. The swing up control problem for the acrobot. *IEEE Control Systems Magazine*, 15:49–55, 1995.

[47] K. O. Stanley and R. Mikkulainen. Evolving neural networks through augmenting topologies. *Evolutionary Computation*, 10:99–127, 2002.

[48] R. Sutton and A. Barto. *Reinforcement Learning: An Introduction*. MIT Press, 1998.

[49] I. Szita and A. Lőrincz. Learning tetris using the noisy cross-entropy method. *Neural Computation*, 18(12):2936–2941, 2006.

[50] I. Szita and A. Lőrincz. Learning to play using low-complexity rule-based policies. illustrations through ms. pac-man. *Journal of Artificial Intelligence Research*, 30:659–684, 2007.

[51] S. Whiteson, M. E. Taylor, and P. Stone. Critical factors in the empirical performance of temporal difference and evolutionary methods for reinforcement learning. *Autonomous Agents and Multi-Agent Systems*, 21(1):1–35, 2010.

[52] S. W. Yoon, A. Fern, and R. Givan. Learning heuristic functions from relaxed plans. In *International Conference on Automated Planning and Scheduling (ICAPS'06)*, pages 162–171, 2006.